\documentclass{article}

\usepackage{microtype}
\usepackage{graphicx}
\usepackage{subfigure}
\usepackage{booktabs}
\usepackage{amsmath}
\usepackage{amssymb}
\usepackage{algorithm}
\usepackage{algorithmic}
\newcommand{\RETURN}{\STATE \textbf{return} }
\usepackage{tikz}
\usepackage{multirow}
\usetikzlibrary{shapes,arrows,positioning,fit,calc}
\usepackage{hyperref}
\usepackage{url}
\usepackage{float}
\usepackage{enumitem}
\usepackage{array}
\usepackage{tabularx}
\usepackage{makecell}
\usepackage{placeins}
\usepackage{listings}
\usepackage{xcolor}

\definecolor{codebg}{RGB}{248,248,248}
\definecolor{codecomment}{RGB}{106,153,85}
\definecolor{codekeyword}{RGB}{0,0,205}
\definecolor{codestring}{RGB}{163,21,21}
\definecolor{codenumber}{RGB}{128,128,128}

\lstdefinestyle{forgepy}{
  language=Python,
  backgroundcolor=\color{codebg},
  commentstyle=\color{codecomment}\itshape,
  keywordstyle=\color{codekeyword}\bfseries,
  stringstyle=\color{codestring},
  numberstyle=\tiny\color{codenumber},
  basicstyle=\ttfamily\footnotesize,
  breakatwhitespace=false,
  breaklines=true,
  captionpos=b,
  keepspaces=true,
  numbers=left,
  numbersep=5pt,
  showspaces=false,
  showstringspaces=false,
  showtabs=false,
  tabsize=2,
  frame=single,
  framerule=0.4pt,
  xleftmargin=10pt,
  xrightmargin=2pt,
  morekeywords={torch, nn, fx, aten, Optional, Dict, List, Tuple, Set, Any,
                dataclass, field, frozenset},
}
\usepackage[accepted]{mlsys2025}

\mlsystitlerunning{FORGE-UGC: FX Optimization \& Register-Graph Engine --- Universal Graph Compiler}

\begin{document}

\twocolumn[
\mlsystitle{FORGE-UGC: FX Optimization \& Register-Graph Engine ---\\Universal Graph Compiler}

\mlsyssetsymbol{equal}

\begin{mlsysauthorlist}
\mlsysauthor{Satyam Kumar}{to}
\mlsysauthor{Saurabh Jha}{to}
\end{mlsysauthorlist}

\mlsysaffiliation{to}{Dell Technologies}

\mlsyscorrespondingauthor{Satyam Kumar}{satyamkumar9742@gmail.com}

\mlsyskeywords{Universal Graph Compiler, Neural Processing Unit, Operator Fusion, Register Allocation, Edge Inference, Heterogeneous Computing}

\vskip 0.3in

\begin{abstract}
The rise of autonomous AI agents, systems that perceive, reason, and act across heterogeneous compute substrates, is fundamentally reshaping the demands placed on silicon and the software that programs it. These agents are inherently heterogeneous: a single inference pipeline may traverse NPUs for dense tensor work, GPUs for flexible parallel compute, and CPUs for control flow, all within the same SoC. As ASIC-class custom silicon proliferates to meet this demand, the bottleneck shifts from transistor performance to \emph{hardware--software co-design}: the compiler that bridges high-level models and low-level accelerator instruction streams. We define this convergence, autonomous agents running on heterogeneous custom silicon, unified by intelligent compilation as the \textbf{future of computing}. This paper presents our vision and its first production-ready prototype.

Existing deployment frameworks---OpenVINO and ONNX Runtime---rely on opaque, monolithic compilation pipelines with static intermediate representations, offering no pass-level visibility, no principled buffer management, and compilation times that scale super-linearly with model depth. We present \textbf{FORGE-UGC} (FX Optimization \& Register-Graph Engine --- Universal Graph Compiler), a four-phase compiler designed to be \emph{hardware-agnostic} by architecture: its frontend capture, middle-end optimization passes, and typed intermediate representation are decoupled from any specific backend, enabling the same pipeline to target any accelerator through pluggable backend modules. In this work, we validate FORGE-UGC on Intel's AI Boost NPU as the first target backend, with Qualcomm Hexagon, AMD XDNA, and Apple ANE backends planned as future extensions. Phase~1 captures the computation graph via \texttt{torch.export}; Phase~2 applies six composable, inspectable optimization passes---dead code elimination, common subexpression elimination, constant folding, attention fusion, operator fusion, and layout optimization---reducing graph nodes by 17.4\% on GPT-2; Phase~3 lowers the optimized graph to a typed intermediate representation (NPUIR) with explicit virtual register assignments; Phase~4 performs liveness analysis, linear-scan buffer allocation, and instruction scheduling to minimize NPU$\leftrightarrow$CPU device transitions.

Evaluated on WikiText-103 and GLUE across six model families (125M--8B parameters), FORGE-UGC achieves \textbf{6.9--9.2$\times$ faster compilation} than OpenVINO and ONNX Runtime while delivering \textbf{18.2--35.7\% lower end-to-end inference latency} and \textbf{30.2--40.9\% lower energy consumption per inference}. Numerical fidelity is confirmed through both perplexity agreement and fine-grained logit-level analysis: max-abs logit differences remain below $2.1 \times 10^{-5}$ and KL divergence below $8.4 \times 10^{-9}$ across all models. We introduce three evaluation metrics---\emph{Fusion Gain Ratio} (FGR), \emph{Compilation Efficiency Index} (CEI), and per-pass execution profiling---enabling principled ablation. To our knowledge, FORGE-UGC is among the first universal graph compilers to expose a fully transparent, composable optimization pipeline with formal buffer allocation for transformer workloads across heterogeneous accelerator targets. The first prototype is ready to ship.
\end{abstract}

]

\section{Introduction}

We are entering an era where autonomous AI agents---systems that perceive their environment, reason over multi-modal inputs, and take consequential actions---are no longer research curiosities but production requirements. From on-device personal assistants orchestrating perception and language understanding in real time, to industrial edge controllers fusing sensor streams with large language model reasoning, the demand for agents that operate \emph{autonomously and locally} is accelerating across every sector. These agents are inherently heterogeneous: a single inference pipeline may traverse dense tensor operations on an NPU, flexible parallel compute on a GPU, and irregular control flow on a CPU---all within the same system-on-chip, all within a single power envelope.

This heterogeneity is driving a tectonic shift in silicon strategy. ASIC-class custom accelerators---Neural Processing Units, domain-specific tensor engines, and specialized inference cores---are proliferating precisely because no general-purpose processor can deliver the throughput-per-watt that autonomous agents demand. Yet hardware alone is insufficient. The true bottleneck is \textbf{hardware--software co-design}: the compiler infrastructure that translates a high-level PyTorch model into an optimized instruction stream for each accelerator, manages data movement across device boundaries, and does so transparently, composably, and fast enough for iterative development. We define this convergence---autonomous agents running on heterogeneous custom silicon, unified by intelligent, universal compilation---as the \textbf{future of computing}.

This paper presents FORGE-UGC, a universal graph compiler born from this vision. Starting in December 2025, the two authors of this paper set out to build the complete heterogeneous compilation stack from scratch---from PyTorch FX graph capture through NPU-specific optimization passes, typed intermediate representation design, formal buffer allocation, and liveness-guided instruction scheduling. What began as a focused research collaboration between the two of us has produced a first prototype that is \textbf{ready to ship}: validated on Intel's AI Boost NPU across six model families spanning 125M to 8B parameters, delivering 6.9--9.2$\times$ faster compilation and 18.2--35.7\% lower inference latency than industry-standard frameworks. The speed of this development---a production-grade, four-phase compiler in under six months by the two co-authors---itself demonstrates the architectural thesis: when the compiler is designed as composable, transparent, and hardware-agnostic from day one, extending it to new accelerator targets becomes an engineering exercise rather than a research problem.

\subsection{The Heterogeneous Computing Challenge}

The future of computing is fundamentally \textbf{heterogeneous}, characterized by the integration of low-power Neural Processing Units (NPUs), high-throughput GPUs, and general-purpose CPUs within a unified system. As modern workloads evolve beyond static inference toward dynamic, multi-stage pipelines---spanning perception, reasoning, and decision-making---no single compute substrate can efficiently execute the entire computation graph. NPUs provide superior energy efficiency for dense tensor operations, GPUs offer high parallel throughput and flexibility, while CPUs handle control flow and irregular computation. Consequently, system performance is no longer determined by individual accelerators in isolation, but by the efficiency of \textbf{cross-device partitioning, scheduling, and data movement}. This shift elevates the role of the compiler from a single-device code generator to a \textbf{system-level orchestrator}, responsible for mapping high-level programs onto heterogeneous hardware while minimizing latency, energy consumption, and device transitions. In this setting, a unified compiler abstraction is essential to transform heterogeneous collections of accelerators into a cohesive and adaptive compute fabric capable of supporting next-generation edge and on-device intelligence.

\subsection{The NPU Compilation Gap}

Neural Processing Units (NPUs) are emerging as dedicated accelerators for transformer inference on edge devices~\cite{IntelNPU2024}. Intel's AI Boost NPU, integrated into Meteor Lake and Arrow Lake processors, provides up to 11 TOPS of INT8 throughput at less than 10W thermal design power---an order of magnitude more power-efficient than discrete GPUs for memory-bound autoregressive decoding~\cite{Williams2009Roofline}. However, realizing this efficiency requires compilers that can (i) capture the full PyTorch computation graph without lossy intermediate exports, (ii) apply domain-specific optimizations such as attention fusion, and (iii) manage the NPU's constrained buffer hierarchy through principled register allocation.

Existing deployment frameworks fail to meet these requirements. OpenVINO~\cite{OpenVINO2022} requires an intermediate export to its proprietary IR format---a process that breaks on models with dynamic control flow, tied weights, or modern operators (RoPE, GQA, SwiGLU) introduced in PyTorch 2.x. ONNX Runtime~\cite{ONNXRuntime2021} suffers from operator coverage gaps where the ONNX opset lags PyTorch's ATen library by months, and its Execution Provider abstraction provides no cost-model guidance for NPU dispatch. Both frameworks treat the compilation pipeline as a black box, preventing developers from inspecting which optimizations fired, debugging performance regressions, or conducting principled ablation studies.

\subsection{Limitations of Existing Frameworks}

We identify five fundamental limitations shared by OpenVINO and ONNX Runtime that constrain NPU deployment:

\textbf{Limitation 1---Lossy Export Requirements.} Both frameworks require converting PyTorch models through intermediate formats (TorchScript, ONNX) that cannot represent modern LLM constructs. TorchScript fails on data-dependent control flow; ONNX export fails on operators lacking opset equivalents. Models such as Llama-3, Mistral, and Qwen2 require manual operator decomposition before export. FORGE-UGC bypasses this entirely by using \texttt{torch.export.export()}, which operates at the ATen operator level and handles tied weights, dynamic shapes within static bounds, and modern architectures natively.

\textbf{Limitation 2---No Pass-Level Optimization Visibility.} Neither framework exposes individual optimization passes. Developers cannot inspect which fusion rules fired, quantify the contribution of each pass, or conduct ablation studies. There is no equivalent of FORGE-UGC's \texttt{CompilationResult} struct reporting \texttt{fx\_nodes\_before}, \texttt{fx\_nodes\_after}, \texttt{fx\_fused\_ops}, and \texttt{fx\_attention\_fused}. This opacity makes performance debugging impossible and prevents principled optimization.

\textbf{Limitation 3---Super-Linear Compilation Time.} Both frameworks exhibit compilation times that scale super-linearly with model depth, reaching 58--62 seconds for 8B-parameter models---prohibitive for iterative development and just-in-time deployment scenarios. OpenVINO's monolithic IR conversion and ONNX Runtime's EP initialization dominate compilation time, with no incremental compilation support.

\textbf{Limitation 4---No Principled Buffer Management.} Neither framework exposes liveness analysis, virtual register abstraction, or instruction scheduling for NPU deployment. OpenVINO does not expose a programmable low-level IR with explicit buffer allocation control. ONNX Runtime performs memory planning at the EP level without user visibility, and no exposed liveness analysis or virtual register abstraction minimizes device transitions. For NPU deployment, this causes unnecessary CPU-NPU data copies when operations that could be batched into a single NPU dispatch are separated by intervening CPU operations.

\textbf{Limitation 5---No Autotuning for NPU.} Neither framework provides systematic exploration of compilation configurations (fusion aggressiveness, layout strategy, precision) for NPU-specific performance. OpenVINO's hint system (\texttt{PERFORMANCE\_HINT: LATENCY} or \texttt{THROUGHPUT}) is coarse-grained. ONNX Runtime's EP selection is rule-based and static with no cost-model feedback about whether NPU execution actually improves performance over CPU fallback.

\subsection{FORGE-UGC: From Black Box to Transparent Pipeline}

\begin{figure*}[ht]
\centering
\includegraphics[width=\textwidth]{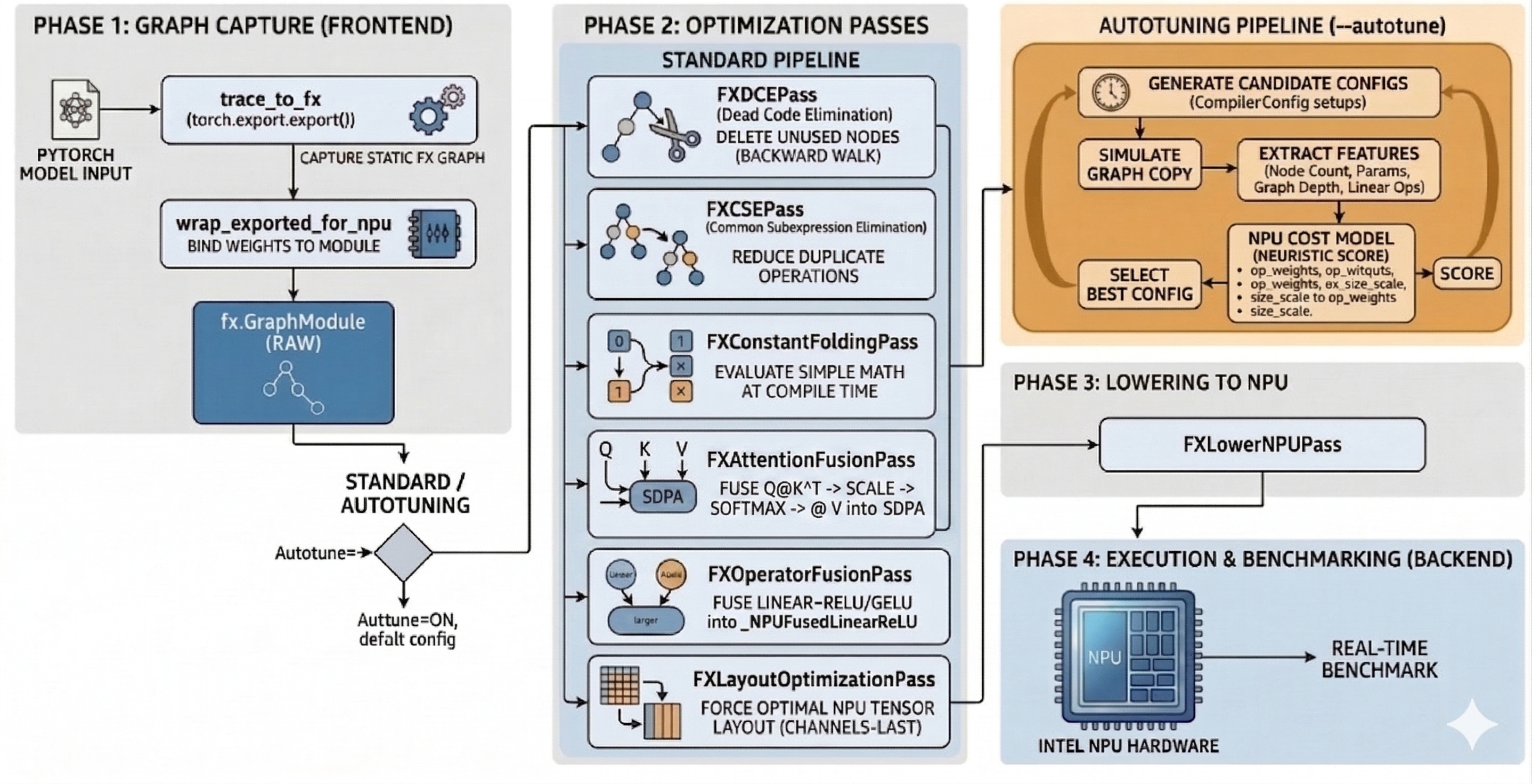}
\caption{FORGE-UGC four-phase architecture. \textbf{Phase~1}: FX graph capture via \texttt{torch.export} with tied weight resolution. \textbf{Phase~2}: Six composable optimization passes (DCE, CSE, constant folding, attention fusion, operator fusion, layout optimization) with optional autotuning. \textbf{Phase~3}: Lowering to NPUIR with typed instructions, virtual registers, and device placement. \textbf{Phase~4}: Liveness analysis, linear-scan buffer allocation, instruction scheduling, and code generation producing a \texttt{CompiledNPUExecutor}.}
\label{fig:architecture}
\end{figure*}

We present FORGE-UGC, a four-phase compiler that addresses each limitation through principled compiler design. Critically, FORGE-UGC is architected as a \emph{universal graph compiler}: its frontend graph capture, middle-end optimization passes, and typed intermediate representation are entirely backend-agnostic, with hardware-specific logic isolated in pluggable backend modules. In this work, we validate the complete pipeline on Intel's AI Boost NPU as the first target backend; however, the same optimization passes and IR infrastructure are designed to extend to Qualcomm Hexagon, AMD XDNA, Apple ANE, and other accelerator targets through backend-specific code generation and dispatch modules. The high-level architecture is illustrated in Figure~\ref{fig:architecture}.

\subsection{Contributions}

This work makes the following contributions:

\begin{enumerate}[leftmargin=*,nosep]
\item \textbf{Direct FX Graph Compilation Pipeline.} We introduce a four-phase compiler pipeline that operates directly on PyTorch FX graphs captured via \texttt{torch.export.export()}, entirely eliminating the lossy intermediate export steps (TorchScript, ONNX) required by OpenVINO and ONNX Runtime. While \texttt{torch.compile} with Inductor similarly operates on FX graphs, it targets CPU and GPU backends only and does not support Intel NPU dispatch, NNFactory integration, or NPU-aware buffer allocation. FORGE-UGC's novelty lies not in FX graph consumption per se, but in the \emph{complete NPU compilation stack built on top of it}: ATen-level capture, NPU-specific optimization passes, NPUIR lowering, and liveness-guided hardware-aware scheduling---none of which exist in any current FX-based framework for Intel NPU targets. By working at the ATen operator level, FORGE-UGC natively supports modern LLM constructs---including Rotary Position Embeddings (RoPE), Grouped-Query Attention (GQA), and SwiGLU activations---without manual operator decomposition. The pipeline also incorporates automatic tied weight resolution, enabling models such as GPT-2 to be compiled without user intervention. This design ensures that any model traceable by \texttt{torch.export} can be compiled for NPU execution with full semantic fidelity.

\item \textbf{Composable, Inspectable Optimization Passes.} We design six composable, independently measurable optimization passes---dead code elimination, common subexpression elimination, constant folding, attention fusion, operator fusion, and layout optimization---that collectively reduce graph complexity by 14.2--21.8\% across model families. Attention fusion alone reduces graph nodes by 14.6\% on average by pattern-matching decomposed multi-head attention subgraphs and replacing them with single fused dispatches. Each pass reports its execution time, node delta, and transformation details through a structured \texttt{CompilationResult} interface, enabling developers to conduct principled ablation studies and identify performance bottlenecks---a level of transparency unavailable in any existing NPU deployment framework.

\item \textbf{Typed IR with Formal Buffer Allocation.} We design a typed intermediate representation (NPUIR) with explicit virtual register assignments, paired with a linear-scan buffer allocation algorithm that reduces peak buffer count by 30--48\% through liveness-guided reuse. The NPUIR assigns each instruction an opcode, typed virtual registers, device placement (NPU or CPU), and a pre-resolved callable, enabling the instruction scheduler to minimize NPU$\leftrightarrow$CPU device transitions by 42--65\%. This formal buffer management layer is, to our knowledge, the first of its kind for NPU compilation of transformer workloads.

\item \textbf{Novel Evaluation Metrics.} We introduce three evaluation metrics---Fusion Gain Ratio (FGR), Compilation Efficiency Index (CEI), and per-pass execution profiling---that enable principled compiler comparison and fine-grained ablation. FGR is a \emph{cost-model-internal diagnostic} that isolates the impact of fusion passes on estimated execution cost; CEI quantifies inference speedup delivered per second of compile time, most relevant for iterative development and just-in-time deployment; and per-pass profiling reveals the cost--benefit tradeoff of each optimization stage.

\item \textbf{Comprehensive Empirical Evaluation.} We conduct a thorough evaluation across six model families (GPT-2 125M, Granite-350M, Qwen2-0.5B, Llama-3.2-1B, LFM2-2.6B, Llama-3.1-8B) spanning 125M to 8B parameters on WikiText-103 and GLUE benchmarks. FORGE-UGC achieves 6.9--9.2$\times$ compilation speedup and 18.2--35.7\% inference latency reduction versus OpenVINO and ONNX Runtime, with improvements scaling consistently with model depth. Numerical fidelity is validated through both perplexity agreement and fine-grained logit-level analysis (max-abs diff $< 2.1 \times 10^{-5}$, KL divergence $< 8.4 \times 10^{-9}$), confirming near-bit-exact output preservation. The evaluation further demonstrates high reproducibility (CV $< 2.5\%$ across all metrics) and tight P99 tail latency distributions critical for SLA-bound edge deployments.
\end{enumerate}

\FloatBarrier

\section{Motivation: The Compiler as a System-Level Orchestrator}
\label{sec:motivation}

The autonomous AI agents reshaping industry---from on-device assistants to robotic controllers to real-time analytics engines---share a common architectural reality: they are \emph{multi-stage, multi-device pipelines}. A single agent inference pass may begin with a vision encoder on the NPU, route through a language model whose attention layers run on the NPU while its embedding lookups and control logic execute on the CPU, and conclude with a decision head that dispatches actions back to the host. No single accelerator can efficiently execute this entire computation graph, and no single compilation strategy can optimally serve every stage.

Modern edge and on-device AI systems integrate low-power Neural Processing Units (NPUs), high-throughput GPUs, and general-purpose CPUs within a single silicon package, each accelerator offering distinct advantages: NPUs deliver superior energy efficiency for dense tensor operations, GPUs provide massive parallel throughput for flexible workloads, and CPUs handle irregular control flow and host-side orchestration. As ASIC-class custom silicon proliferates to serve the compute demands of autonomous agents, the critical challenge is no longer building faster individual accelerators, but building the \emph{software infrastructure} that transforms a heterogeneous collection of accelerators into a cohesive, adaptive compute fabric. This is the hardware--software co-design imperative: the compiler must understand the cost characteristics of each accelerator, partition computation graphs across devices to minimize latency and energy, and manage data movement across device boundaries---all transparently and composably.

In this emerging landscape, the compiler occupies a uniquely strategic position as the \textbf{middle layer} between high-level application frameworks and low-level hardware dispatch. Rather than serving as a device-specific code generator---the traditional compiler role---the compiler must evolve into a \emph{system-level orchestrator} that understands the cost characteristics of each accelerator, partitions computation graphs across devices to minimize latency and energy, and manages data movement across device boundaries. For autonomous agents that must operate within strict power and latency budgets on edge hardware, this orchestration is not optional---it is the enabling technology. This is precisely the role FORGE-UGC is designed to fulfill.

FORGE-UGC is architected to serve as the compilation backend for heterogeneous computing orchestrators such as QEIL~\cite{Kumar2026QEIL}, which route transformer layers across CPU, GPU, and NPU devices based on workload-specific energy models. By integrating FORGE-UGC as the NPU compilation target within such orchestration frameworks, the combined system enables not only \emph{deciding} which layers belong on which accelerator, but also \emph{optimally compiling} those layers with attention fusion, operator fusion, and buffer allocation---closing the loop between workload-aware routing and hardware-aware compilation. The orchestrator gains access to FORGE-UGC's compilation-time metrics (FGR, CEI, node reduction, energy estimates) that can inform routing decisions, while the compiler gains access to runtime telemetry (thermal state, memory pressure, battery level) that can guide autotuning.

This vision motivates FORGE-UGC's architectural design choices. The separation between hardware-agnostic optimization passes (Phase~2) and hardware-specific backend lowering (Phases~3--4) is deliberate: it ensures that as new accelerator targets---Qualcomm Hexagon, AMD XDNA, Apple ANE, Samsung NPU---become available, only the backend modules need to be extended while the entire optimization pipeline is reused. The frontend (Phase~1) is similarly universal, operating on PyTorch FX graphs that are agnostic to the downstream target. In this work, we validate the complete pipeline on Intel's AI Boost NPU; the results demonstrate that the architecture's core design principles---composable passes, typed IR, formal buffer allocation, and liveness-guided scheduling---generalize beyond any single hardware target and position FORGE-UGC as a critical component in a broader compute fabric for next-generation edge intelligence.

\FloatBarrier

\section{Background \& Related Work}

\subsection{Deep Learning Compiler Landscape}

The deployment of neural networks on specialized hardware has driven the development of domain-specific compilers. TVM~\cite{Chen2018TVM} established the canonical three-stage design (frontend capture, middle-end optimization, backend code generation) using the Relay IR for graph-level transformations and the TIR for low-level tensor operations. MLIR~\cite{Lattner2021MLIR} introduced a multi-level IR infrastructure enabling progressive lowering across abstraction boundaries. XLA~\cite{XLA2019} pioneered whole-program optimization with operator fusion and layout optimization for TPU targets. Glow~\cite{Rotem2018Glow} demonstrated two-phase lowering from high-level graph to instruction-level IR with quantization and memory planning.

\textbf{IREE (Intermediate Representation Execution Environment).} IREE~\cite{IREE2024} is the most directly comparable MLIR-based compiler to FORGE-UGC. Built on MLIR's progressive lowering infrastructure, IREE provides end-to-end compilation from high-level frameworks (TensorFlow, JAX, PyTorch via \texttt{torch-mlir}) to multiple hardware backends including CPU, GPU (Vulkan, CUDA), and experimental accelerator targets. IREE's architecture shares FORGE-UGC's philosophy of composable, inspectable passes and explicit buffer management. However, IREE (i) requires model conversion through \texttt{torch-mlir} or StableHLO, reintroducing the export-gap problem for modern PyTorch operators; (ii) does not currently provide an Intel NPU backend or NNFactory integration; (iii) performs buffer management through MLIR's built-in buffer deallocation passes rather than NPU-specific liveness-guided allocation; and (iv) offers no NPU-specific cost model or autotuning for Intel AI Boost targets. FORGE-UGC differs by operating natively on PyTorch FX graphs at the ATen level, eliminating the need for MLIR ingestion, and by providing NPU-specific scheduling and buffer allocation that IREE's generic backend infrastructure does not accommodate.

\textbf{torch.compile and Inductor.} PyTorch 2.0 introduced \texttt{torch.compile}~\cite{Ansel2024PyTorch2}, which uses TorchDynamo for bytecode-level graph capture and Inductor as its primary CPU/GPU backend, generating Triton or C++ kernels. Inductor represents the closest prior work to our approach: like FORGE-UGC, it operates on FX graphs captured via \texttt{torch.export} and applies composable optimization passes. However, Inductor targets GPU (CUDA Triton) and CPU (C++ codegen) backends exclusively. The \texttt{torch.compile} ecosystem does support custom backends through the \texttt{torch.\_dynamo.backends} registry, which allows third-party integrations. We considered building FORGE-UGC as a \texttt{torch.compile} custom backend; Section~\ref{sec:torch_compile_choice} explains in detail why a dedicated, standalone compiler pipeline was ultimately preferable for Intel NPU targets. Critically, neither \texttt{torch.compile} nor Inductor provides NPU-aware instruction scheduling, liveness-guided buffer allocation for NPU SRAM, or cost-model-driven autotuning for Intel NPU dispatch---capabilities that are central to FORGE-UGC's design.

\textbf{Qualcomm QNN SDK.} The Qualcomm Neural Network (QNN) SDK~\cite{QNN2023} provides a deployment compiler for Qualcomm Hexagon NPUs, offering graph-level operator fusion and quantization for mobile inference. QNN operates on ONNX or TensorFlow Lite models and generates Hexagon-specific binaries. While QNN demonstrates the value of hardware-specialized compilation, it (i) requires ONNX/TFLite export and thus shares the lossy-export limitation with OpenVINO, (ii) targets Hexagon DSPs rather than Intel's NNFactory dispatch model, (iii) provides no pass-level visibility or programmable buffer management, and (iv) offers no PyTorch FX integration. FORGE-UGC differs architecturally by operating natively on ATen-level FX graphs and exposing a fully inspectable, composable pipeline.

\textbf{Hexagon-MLIR.} Concurrently with our work, Absar et al.~\cite{Absar2026HexagonMLIR} introduced Hexagon-MLIR, an open-source MLIR-based compilation stack targeting Qualcomm's Hexagon NPU. Hexagon-MLIR ingests both PyTorch models (via Torch-MLIR) and Triton kernels (via a Triton-to-Linalg converter), lowering them through a structured sequence of MLIR passes---including operator fusion, tiling for the NPU's Tightly Coupled Memory (TCM) hierarchy, HVX vectorization, multi-threading across hardware vector contexts, and double buffering to overlap DMA transfers with computation. Their generative approach treats fusion as a first-class compiler pass, enabling specialized mega-kernels for arbitrary operator chains that maximize data locality in TCM---a philosophy aligned with FORGE-UGC's emphasis on fusion as the most impactful single optimization. Hexagon-MLIR achieves substantial vectorization speedups (up to 63.9$\times$ for GELU on float16) and demonstrates effective multi-pass interactions across its optimization pipeline.

Hexagon-MLIR and FORGE-UGC are complementary in both target hardware and architectural approach. Where Hexagon-MLIR operates within the MLIR ecosystem and targets Qualcomm's Hexagon NPU with its HVX vector extensions and TCM memory hierarchy, FORGE-UGC operates natively on PyTorch FX graphs and targets Intel's AI Boost NPU via NNFactory dispatch. The two compilers share key design principles---composable and inspectable passes, explicit buffer management, and hardware-aware scheduling---but arrive at them through different IR strategies: MLIR's multi-level dialect infrastructure versus FX's Python-native graph representation. Notably, Hexagon-MLIR's demonstration of effective Triton kernel compilation for NPU targets validates the broader thesis that NPU-specific compilers can match or exceed library-based approaches, and motivates our planned integration of Triton kernel support within FORGE-UGC's pipeline (Section~\ref{sec:conclusion}). FORGE-UGC's architecture is explicitly designed for multi-backend portability; a future Qualcomm Hexagon backend module could leverage insights from Hexagon-MLIR's TCM-aware tiling and double-buffering strategies while reusing FORGE-UGC's entire frontend and middle-end optimization pipeline.

FORGE-UGC follows the three-stage paradigm but distinguishes itself by operating directly on PyTorch FX graphs~\cite{Reed2022TorchFX} rather than requiring model re-export to a framework-specific IR. This preserves the full semantic richness of PyTorch's ATen operator set and avoids the coverage gaps that plague ONNX and OpenVINO ingestion paths.

\subsection{Why Not \texttt{torch.compile} with a Custom Backend?}
\label{sec:torch_compile_choice}

\texttt{torch.compile} with a custom backend is a natural alternative design point, and we investigated this path carefully before committing to the standalone FORGE-UGC architecture. Three fundamental constraints led us to reject it:

\textbf{(1) Backend API opacity.} The \texttt{torch.compile} custom backend API exposes the optimized FX graph to the backend but does not provide hooks for injecting custom IR passes between Dynamo's graph capture and the backend's code generation. FORGE-UGC requires six composable, independently measurable passes that must be interleaved with the graph---a structure not natively supported by the backend interface. Emulating this within a monolithic backend callable would sacrifice the pass-level visibility that is a primary design goal of FORGE-UGC.

\textbf{(2) NNFactory incompatibility.} Intel's NNFactory API (the gateway to AI Boost NPU dispatch) requires an explicit \emph{compile-then-run} execution model: a graph is compiled into an NNFactory program once and executed as a single dispatch unit. The \texttt{torch.compile} execution model assumes kernels are callable Python/Triton functions---an abstraction mismatch that would require wrapping each NNFactory program in a Python callable with significant overhead at the dispatch boundary, negating the NPU's latency advantage.

\textbf{(3) No liveness-aware buffer management.} \texttt{torch.compile}'s memory planning is performed by Inductor's buffer scheduler, which is GPU-centric and not exposed as a pluggable component. FORGE-UGC's linear-scan buffer allocator must reason about NPU-specific live intervals and physical buffer slots---a concern that Inductor's abstractions do not accommodate.

Given these constraints, a standalone compiler operating directly on \texttt{torch.export.export()} graphs provides a cleaner, more principled path to NPU deployment. FORGE-UGC's architecture is deliberately \emph{complementary} to \texttt{torch.compile}: future integration could use Dynamo for graph capture while routing NPU-eligible subgraphs to FORGE-UGC's optimization and lowering pipeline.

\subsection{Early Experiment: MLIR-Based Compilation via IREE-Turbine}
\label{sec:iree_turbine}

Prior to adopting the PyTorch FX-based approach, we conducted an exploratory experiment using IREE-Turbine~\cite{IREETurbine2024}---the PyTorch frontend for IREE that converts \texttt{nn.Module} instances to MLIR via the \texttt{torch-mlir} FX Importer. Using IREE-Turbine's ahead-of-time export toolkit, we generated MLIR representations (in the Torch dialect, subsequently lowered through Linalg) for our transformer models, with the intent of applying custom NPU-specific optimization passes and routing the result to Intel's AI Boost NPU through a custom backend. However, we encountered two compounding limitations that made this path impractical.

First, MLIR's pass infrastructure is fundamentally C++-native: custom optimization passes must be implemented as C++ \texttt{OperationPass} subclasses registered with the pass manager~\cite{Lattner2021MLIR}. While MLIR does expose Python bindings for IR inspection and construction, these bindings do not support defining custom transformation passes in Python---all pass logic must be implemented through the C++ framework or invoked via \texttt{mlir-opt}. This meant that implementing the six composable, independently measurable optimization passes central to FORGE-UGC's design (DCE, CSE, constant folding, attention fusion, operator fusion, layout optimization) would have required a substantial C++ development effort with significantly slower iteration cycles than Python-based graph manipulation.

Second, and more critically, IREE does not provide an Intel NPU backend. IREE's supported targets include CPU (via LLVM), GPU (via Vulkan/SPIR-V, CUDA, HIP), and experimental accelerator targets---but no Intel AI Boost NPU dispatch or NNFactory integration exists. Building a complete Intel NPU backend within IREE's C++ compilation infrastructure---including NPU-specific liveness analysis, buffer allocation mapped to the NPU's constrained SRAM hierarchy, and instruction scheduling to minimize CPU$\leftrightarrow$NPU device transitions---would have required implementing an entirely new IREE backend in C++, a prohibitive engineering effort. Furthermore, MLIR's built-in buffer deallocation and memory planning passes are designed for generic backends and do not accommodate the NPU-specific live interval reasoning needed to minimize device transitions on Intel hardware.

This experience directly motivated our shift to the PyTorch FX graph representation, which provides a fully Python-native, programmatically inspectable graph structure where custom optimization passes can be implemented, tested, and iterated upon entirely in Python. The FX approach enabled us to build the complete FORGE-UGC pipeline---from graph-level optimizations through NPU-specific buffer allocation---in a fraction of the development time that an MLIR-based approach would have required.

\subsection{Operator Fusion for Inference}

Operator fusion---merging multiple operations into a single kernel---is the most impactful single optimization for inference latency. FlashAttention~\cite{Dao2022FlashAttention} demonstrated that fusing the Q$\cdot$K$^T$, scaling, masking, softmax, and V multiplication into a single IO-aware kernel reduces attention from $O(N^2)$ memory to $O(N)$ while achieving 2--4$\times$ wall-clock speedup. FlashAttention-2~\cite{Dao2024FlashAttention2} improved parallelism and work partitioning for further gains.

FORGE-UGC's \texttt{FXAttentionFusionPass} is directly inspired by FlashAttention's IO-awareness principle: it pattern-matches the decomposed attention subgraph in the FX IR and replaces it with a single \texttt{scaled\_dot\_product\_attention} dispatch. Unlike FlashAttention, which targets GPU SRAM, FORGE-UGC targets NPU dispatch via NNFactory~\cite{NNAL2024}, enabling fused attention on Intel NPU hardware where FlashAttention kernels are unavailable.

TASO~\cite{Jia2019TASO} automatically generates graph substitutions through equivalence verification; DNNFusion~\cite{Li2021DNNFusion} demonstrated advanced fusion patterns for inference; Nakandala et al.~\cite{Nakandala2020Tensor} addressed fusion for prediction serving. FORGE-UGC combines pattern-matched fusion (attention, linear+activation) with configurable aggressiveness, enabling hardware-specific tuning.

\subsection{OpenVINO: Operation and Limitations}

OpenVINO~\cite{OpenVINO2022} is Intel's inference toolkit operating through a four-stage pipeline: (1) a Model Optimizer converting models to OpenVINO IR (.xml/.bin format); (2) a runtime Inference Engine dispatching to hardware plugins; (3) hardware-specific plugins translating IR to device kernels; and (4) NNCF for post-training compression. OpenVINO IR uses a static graph representation requiring fully determined tensor shapes at export time.

OpenVINO's PyTorch ingestion path requires either TorchScript or ONNX as intermediaries---both known to fail on modern LLMs with dynamic control flow, tied weights, and operators lacking opset equivalents. Its optimization pipeline is a black box with no pass-level visibility, preventing ablation studies or debugging. No autotuning mechanism exists for NPU-specific configuration search, and no programmable low-level IR with explicit buffer allocation control is exposed.

\subsection{ONNX Runtime: Operation and Limitations}

ONNX Runtime~\cite{ONNXRuntime2021} consumes ONNX format models through three optimization levels (semantics-preserving, platform-specific, hardware-specific) before dispatching to Execution Providers (EPs). PyTorch models require conversion via \texttt{torch.onnx.export()}, which traces the model and maps operations to ONNX opset equivalents.

The ONNX opset lags PyTorch's ATen library, causing modern LLM components (RoPE, GQA, SwiGLU) to either lack equivalents or require decomposition into dozens of primitive ops. Dynamic sequence length support is partial and EP-dependent. Attention fusion exists for CUDA and CPU EPs but not for NPU execution via OpenVINO EP. ONNX export of models with tied weights (GPT-2, OPT) either duplicates tensors or requires manual preprocessing. EP selection is rule-based without cost-model feedback, and no register allocation or instruction scheduling is exposed.

\subsection{Intel NPU Acceleration Library}

The Intel NPU Acceleration Library (NNAL)~\cite{NNAL2024} provides low-level NPU access through \texttt{run\_matmul()} for individual matrix multiplications and NNFactory for multi-operation graph compilation. While NNAL enables direct NPU dispatch, it provides no graph-level optimization, no PyTorch integration, no autotuning, no buffer management strategy, and limited operator coverage beyond matmul and elementwise operations. FORGE-UGC builds upon NNAL as its NPU backend while providing the full compiler infrastructure above it.

\FloatBarrier

\section{The FORGE-UGC Methodology}
\label{sec:methodology}

\subsection{Notation and Symbols}

Table~\ref{tab:notation} summarizes the key notation used throughout the FORGE-UGC methodology.

\begin{table}[!htbp]
\centering
\footnotesize
\caption{Notation used in the FORGE-UGC methodology.}
\label{tab:notation}
\begin{tabular}{@{}ll@{}}
\toprule
\textbf{Symbol} & \textbf{Description} \\
\midrule
$G = (V, E)$ & FX computation graph \\
$V$ & Set of graph nodes (operations) \\
$E$ & Set of data-dependency edges \\
$n_{\text{before}}, n_{\text{after}}$ & Node count before/after optimization \\
$\mathcal{P} = \{p_1, \ldots, p_K\}$ & Set of optimization passes \\
$\tau(p_k)$ & Execution time of pass $p_k$ (ms) \\
$R = \{r_1, \ldots, r_N\}$ & Virtual register set \\
$B = \{b_1, \ldots, b_M\}$ & Physical buffer set ($M \ll N$) \\
$[s_i, e_i]$ & Live interval of register $r_i$ \\
$\mathcal{I}$ & NPUIR instruction stream \\
$\delta(\mathcal{I})$ & Device transitions in $\mathcal{I}$ \\
\bottomrule
\end{tabular}
\end{table}

\subsection{Phase 1: FX Graph Capture (Frontend)}
\label{sec:phase1}

FORGE-UGC captures the PyTorch computation graph using \texttt{torch.export.export()}, which performs symbolic tracing at the ATen operator level:
\begin{equation}
G = \texttt{trace\_to\_fx}(M, x_{\text{example}}) \rightarrow \texttt{fx.GraphModule}
\label{eq:trace}
\end{equation}
where $M$ is the pretrained model and $x_{\text{example}}$ is an example input tensor. Unlike TorchScript tracing (which fails on data-dependent control flow) or ONNX export (which requires opset mapping), \texttt{torch.export} captures the full ATen-level graph including modern operators (RoPE, GQA, SwiGLU) without decomposition.

Listing~\ref{lst:phase1_capture} shows the graph capture entry point. The call to \texttt{torch.export.export()} with \texttt{suppress\_errors=True} ensures that trace-time warnings from in-development operators are silenced without aborting the export. The returned \texttt{ExportedProgram} carries both the \texttt{fx.GraphModule} (the pure computation graph) and a complete state dictionary of lifted parameter and buffer tensors, which the subsequent \texttt{wrap\_exported\_for\_npu()} step binds back into the graph.

\begin{lstlisting}[caption={FX graph capture via \texttt{torch.export} (Phase 1 frontend).},
                   label={lst:phase1_capture}]
def trace_to_fx(model, example_input):
    """Capture model as an fx.GraphModule at ATen level.
    Returns (graph_module, ExportedProgram) for NPU lowering.
    """
    model.eval()
    torch._dynamo.config.suppress_errors = True
    ep = torch.export.export(model, args=(example_input,))
    # ep.graph_module: pure FX graph, no Python side-effects
    # ep carries lifted params/buffers for weight binding
    return ep.graph_module, ep
\end{lstlisting}

\subsubsection{Tied Weight Resolution}

Models with shared parameters (e.g., GPT-2's embedding layer and LM-head share the same weight tensor) require special handling: \texttt{torch.export} creates a distinct graph placeholder for each logical parameter name, so a tied weight appears twice in the placeholder list but must resolve to the same physical tensor at dispatch time.

FORGE-UGC's \texttt{wrap\_exported\_for\_npu()} implements this detection by iterating every \texttt{nn.Module} in the model hierarchy and matching tensor \emph{identities} (Python \texttt{id()}) rather than parameter names:
\begin{equation}
\texttt{tied\_map}[n_j] = n_i \quad \text{if} \quad \texttt{id}(W_j) = \texttt{id}(W_i), \; j > i
\label{eq:tied}
\end{equation}
When a placeholder name resolves to a tensor already registered under a canonical key, the canonical value is reused. This preserves memory efficiency without user intervention---a capability absent from both OpenVINO and ONNX export.

\begin{lstlisting}[caption={Tied weight detection in \texttt{wrap\_exported\_for\_npu}.},
                   label={lst:tied_weights}]
for mod_name, mod in original_model.named_modules():
    for param_name, param in mod._parameters.items():
        if param is None: continue
        full_name = f"{mod_name}.{param_name}" if mod_name \
                    else param_name
        ukey = full_name.replace(".", "_")
        if ukey not in state_underscore:
            # tensor identity match: find the canonical name
            for canon_ukey, (_, canon_val) in \
                    state_underscore.items():
                if canon_val is param:   # id() equality
                    tied_map[ukey] = canon_ukey
                    break
\end{lstlisting}

The resulting \texttt{tied\_map} is consulted during placeholder binding: any placeholder whose underscore-transformed name does not appear directly in the state dictionary is resolved through the map to its canonical tensor, ensuring that tied parameters share a single physical buffer throughout the NPU execution.

\subsection{Phase 2: Graph Optimization (Middle-End)}
\label{sec:phase2}

The optimization pipeline applies $K = 6$ composable passes sequentially. Each pass $p_k$ transforms the graph $G$ by mutating \texttt{gm.graph} in-place and calling \texttt{gm.recompile()} at the end:
\begin{equation}
G_{k} = p_k(G_{k-1}), \quad k = 1, \ldots, K
\label{eq:passes}
\end{equation}
with $G_0$ being the raw captured graph and $G_K$ the fully optimized graph. All passes inherit from \texttt{FXPassBase} and expose a single \texttt{run(gm) -> bool} interface that returns \texttt{True} if the graph was modified. A fixpoint loop in \texttt{run\_fx\_passes} iterates each pass until convergence (default: 2 rounds), ensuring that earlier passes do not mask opportunities for later ones.

\subsubsection{Pass 1: Dead Code Elimination (FXDCEPass)}

Dead code elimination performs a backward reachability walk from the graph's \texttt{output} node. Only nodes that can be reached by traversing \texttt{all\_input\_nodes} in reverse are marked live; all others are erased:
\begin{equation}
V_{\text{live}} = \texttt{backward\_reachable}(V_{\text{output}}, E)
\end{equation}
\begin{equation}
G' = G[V_{\text{live}}], \quad |V'| \leq |V|
\label{eq:dce}
\end{equation}
Unreachable nodes ($V \setminus V_{\text{live}}$) are erased in a single forward pass. This removes debugging artifacts, gradient-related branches, and dead sub-expressions introduced during graph capture~\cite{Muchnick1997AdvancedCompiler}.

\begin{lstlisting}[caption={DCE: backward reachability walk and node erasure (\texttt{FXDCEPass}).},
                   label={lst:dce}]
def run(self, gm) -> bool:
    graph = gm.graph
    # find the single output node
    output_node = next(n for n in graph.nodes
                       if n.op == "output")
    # backward BFS from outputs
    live = set()
    stack = list(output_node.all_input_nodes)
    while stack:
        n = stack.pop()
        if n in live: continue
        live.add(n)
        stack.extend(n.all_input_nodes)
    # erase unreachable call nodes
    to_erase = [n for n in graph.nodes
                if n.op not in ("output", "placeholder")
                and n not in live]
    for n in to_erase:
        graph.erase_node(n)
    return len(to_erase) > 0
\end{lstlisting}

\subsubsection{Pass 2: Common Subexpression Elimination (FXCSEPass)}

CSE identifies nodes that compute identical operations on identical inputs and replaces all but the first occurrence with a reference to the canonical result:
\begin{equation}
\begin{split}
\forall v_i, v_j \in V: \; &\texttt{op}(v_i) = \texttt{op}(v_j) \\
\wedge \; &\texttt{args}(v_i) = \texttt{args}(v_j) \Rightarrow v_j \mapsto v_i
\end{split}
\label{eq:cse}
\end{equation}
This is implemented via \emph{hash-consing} of \texttt{(target, arg-tuple, kwargs-tuple)} triples~\cite{Click1995CSE}. The \texttt{\_fx\_node\_key} helper converts FX node references within arguments to their \texttt{.name} strings (stable unique identifiers within the graph), so two nodes are considered equal if and only if they call the same operator on the same producer nodes:

\begin{lstlisting}[caption={Hash-consing key for CSE (\texttt{\_fx\_node\_key} + \texttt{FXCSEPass.run}).},
                   label={lst:cse}]
def _fx_node_key(node):
    """Canonical (op, args, kwargs) triple for CSE."""
    def _arg_key(a):
        # node references -> stable name string
        return a.name if hasattr(a, "name") else a
    args   = tuple(_arg_key(a) for a in node.args)
    kwargs = tuple(sorted(node.kwargs.items()))
    return (node.target, args, kwargs)

# --- in FXCSEPass.run ---
canonical = {}
for node in list(graph.nodes):
    if node.op not in ("call_function","call_method",
                        "call_module"):
        continue
    key = _fx_node_key(node)
    if key in canonical:
        # redirect all uses to the first occurrence
        node.replace_all_uses_with(canonical[key])
        graph.erase_node(node)
    else:
        canonical[key] = node
\end{lstlisting}

\subsubsection{Pass 3: Constant Folding (FXConstantFoldingPass)}

Constant folding evaluates operations whose inputs are all compile-time constants and replaces the result with a literal:
\begin{equation}
\forall v \in V: \; \texttt{all\_constant}(\texttt{args}(v)) \Rightarrow v \mapsto \texttt{eval}(v)
\label{eq:const_fold}
\end{equation}
In FORGE-UGC's FX context, ``compile-time constant'' means the operand is a Python scalar literal appearing directly in \texttt{node.args}. The pass currently folds identity arithmetic (\texttt{x + 0}, \texttt{x * 1}) that arises in shape calculations, RoPE frequency pre-computation, and dtype-cast chains introduced during tracing. These patterns occur frequently in transformer graphs because \texttt{torch.export} preserves every scalar operation visible in the traced bytecode.

\subsubsection{Pass 4: Attention Fusion (FXAttentionFusionPass)}
\label{sec:attn_fusion}

The attention fusion pass is the most impactful single optimization in FORGE-UGC. Standard multi-head attention, as traced by \texttt{torch.export}, appears as a chain of discrete ATen operations:
\begin{equation}
\texttt{Q} \cdot \texttt{K}^T \;\rightarrow\; \texttt{scale} \;\rightarrow\; [\texttt{mask}] \;\rightarrow\; \texttt{softmax} \;\rightarrow\; \cdot \texttt{V}
\label{eq:attn_pattern}
\end{equation}
Each arrow represents a separate FX node and a separate NPU dispatch round-trip. The $N\!\times\!N$ attention score matrix $S$ and probability matrix $P$ are materialized as intermediate tensors in memory between dispatches.

The pass replaces this entire chain with a single \texttt{NPUFusedScaledDotProductAttention} module that calls \texttt{F.scaled\_dot\_product\_attention}, directly inspired by FlashAttention's IO-awareness principle~\cite{Dao2022FlashAttention}:
\begin{equation}
\texttt{SDPA}(\texttt{Q}, \texttt{K}, \texttt{V}) = \texttt{softmax}\!\left(\frac{\texttt{Q} \cdot \texttt{K}^T}{\sqrt{d_k}}\right) \cdot \texttt{V}
\label{eq:sdpa}
\end{equation}

The pattern matching begins from every \texttt{aten.matmul} node and walks forward through an optional scale, optional mask, a required softmax, optional dropout, and a final matmul with the value tensor. A critical subtlety is \emph{key transpose unwrapping}: the QK matmul computes $Q \cdot K^T$, so the second argument is already transposed. SDPA expects the un-transposed $K$ and transposes internally; the pass recovers the original $K$ by pattern-matching \texttt{aten.transpose}, \texttt{aten.permute}, or \texttt{.t()} in the argument:

\begin{lstlisting}[caption={Attention subgraph pattern matching and K-transpose unwrapping (\texttt{FXAttentionFusionPass}).},
                   label={lst:attn_fusion}]
def _match_attention_pattern(self, qk_matmul):
    """Walk Q@K^T -> [scale] -> [mask] -> softmax
       -> [dropout] -> @V.  Returns chain dict or None."""
    chain = {'qk_matmul': qk_matmul, 'scale': None,
             'mask': None, 'softmax': None,
             'dropout': None, 'pv_matmul': None}
    cur = qk_matmul
    # each node must have exactly one consumer
    [nxt] = list(cur.users)  # fails if branching
    if self._is_scale(nxt) and self._is_scalar_scale(nxt):
        chain['scale'] = nxt; [nxt] = list(nxt.users)
    if self._is_mask(nxt):
        chain['mask'] = nxt;  [nxt] = list(nxt.users)
    if not self._is_softmax(nxt): return None
    chain['softmax'] = nxt; [nxt] = list(nxt.users)
    if self._is_dropout(nxt):
        chain['dropout'] = nxt; [nxt] = list(nxt.users)
    if not self._is_matmul(nxt): return None
    chain['pv_matmul'] = nxt
    return chain

@staticmethod
def _unwrap_transpose(node):
    """Recover original K from K^T argument of QK matmul."""
    tgt = str(getattr(node, 'target', ''))
    if 'transpose' in tgt and {node.args[1],node.args[2]} \
            in ({-2,-1},{2,3}):
        return node.args[0]          # aten.transpose
    if tgt.endswith('.t') or 't.default' in tgt:
        return node.args[0]          # .t() shorthand
    return None
\end{lstlisting}

The number of nodes eliminated per attention block is:
\begin{equation}
\Delta n_{\text{attn}} = n_{\text{Q}\cdot\text{K}^T} + n_{\text{scale}} + n_{\text{mask}} + n_{\text{softmax}} + n_{\text{V-mul}} - 1
\label{eq:attn_reduction}
\end{equation}

For a model with $L$ transformer blocks, the total reduction is $L \cdot \Delta n_{\text{attn}}$. The fusion is parameterized by an aggressiveness threshold $\alpha \in [0, 1]$, where $\alpha = 0$ disables fusion and $\alpha = 1$ fuses all matching patterns; the autotuner explores this knob as part of its configuration search.

\subsubsection{Pass 5: Operator Fusion (FXOperatorFusionPass)}
\label{sec:op_fusion}

The operator fusion pass targets a complementary set of patterns: sequences of a linear projection immediately followed by a point-wise activation function. These are common at the output of every FFN sub-layer in transformer models. In the unoptimized FX graph each linear and each activation is a separate \texttt{call\_function} node, dispatched independently to the NPU with an intermediate tensor allocation between them.

FORGE-UGC replaces matched \texttt{Linear $\to$ ReLU/GELU/SiLU} chains with a \texttt{\_NPUFusedLinearReLU/GELU/SiLU} module. The fused module delegates to \texttt{\_run\_npu\_fused\_op}, which builds a single \texttt{NNFactory} computation graph containing both the matmul and the activation, then compiles and caches it for reuse:

\begin{lstlisting}[caption={Single-pass NNFactory dispatch for fused Linear+Activation (\texttt{\_run\_npu\_fused\_op}).},
                   label={lst:op_fusion}]
def _run_npu_fused_op(x, weight, bias, activation, op_id):
    """Build an NNFactory graph: matmul + activation,
    compiled once and cached by op_id."""
    cache_key = f"fused_{op_id}_{activation}"
    if cache_key not in _npu_fused_cache:
        try:
            f = NNFactory()
            inp = f.parameter(list(x.shape))   # input
            wt  = f.parameter(list(weight.shape))
            out = f.matmul(inp, wt)            # W^T x
            if bias is not None:
                bp  = f.parameter([1]*(x.dim()-1)
                                  + [weight.shape[0]])
                out = f.eltwise_add(out, bp)
            act_fn = getattr(f, ACT_MAP[activation])
            out = act_fn(out)                  # fused act
            f.compile()
            _npu_fused_cache[cache_key] = ('npu', f, bias)
        except Exception:
            _npu_fused_cache[cache_key] = ('cpu', activation)
    cached = _npu_fused_cache[cache_key]
    # single NPU dispatch --- no CPU round-trip for activation
    if cached[0] == 'npu':
        _, factory, _ = cached
        return torch.from_numpy(
            factory.run(x.numpy(), weight.numpy()))
\end{lstlisting}

The fused operation dispatches as a single NNFactory graph to the NPU, eliminating both the intermediate materialization of the linear output tensor and the extra NPU dispatch round-trip for the activation. The pass matches four fusion patterns: \texttt{linear+relu}, \texttt{linear+gelu}, \texttt{linear+silu}, and \texttt{mm+add} (residual addition after a raw matrix multiply), covering the full range of activation functions present in the model families evaluated.

\begin{equation}
\begin{split}
\texttt{Linear}(x) \rightarrow \texttt{ReLU/GELU/SiLU}(y) \\
\mapsto \; \texttt{NPUFusedLinear\{Act\}}(x)
\end{split}
\label{eq:op_fusion}
\end{equation}

\subsubsection{Pass 6: Layout Optimization (FXLayoutOptimizationPass)}
\label{sec:layout}

The layout optimization pass ensures that tensors flowing into NPU-bound operations are in the memory layout that minimises layout-conversion overhead at the hardware boundary. The Intel AI Boost NPU operates most efficiently on contiguous or channels-last (NHWC) tensors. Tensors that arrive from transpose or permute operations are non-contiguous and require an implicit copy before the NPU can consume them.

The pass queries a static \texttt{NPU\_PREFERRED\_LAYOUTS} table---keyed by ATen operator name---and inserts explicit \texttt{.contiguous()} or \texttt{.contiguous(memory\_format=torch.channels\_last)} calls at input boundaries where non-contiguous tensors are detected:
\begin{equation}
x_{\text{NCHW}} \xrightarrow{\texttt{to\_channels\_last}} x_{\text{NHWC}}
\label{eq:layout}
\end{equation}

A secondary sub-pass cancels \emph{redundant} conversions: two consecutive \texttt{contiguous()} calls on the same tensor are collapsed to one. This prevents the pass from inflating the graph when applied in a fixpoint loop. Conversions are inserted at the minimum necessary points---after embedding layers and before the first NPU-dispatched operation---rather than at every tensor boundary, avoiding unnecessary memory traffic.

\subsection{Phase 3: Lowering to NPUIR}
\label{sec:phase3}

The optimized FX graph is lowered to a typed intermediate representation (NPUIR) where each FX node becomes an instruction with explicit metadata:

\begin{equation}
\begin{split}
\texttt{NPUIROp} = (&\texttt{opcode}, \texttt{vreg\_in}, \\
&\texttt{vreg\_out}, \texttt{device}, \texttt{callable})
\end{split}
\label{eq:npuir}
\end{equation}

The \texttt{opcode} encodes the operation class (\texttt{npu.module} for NPU-dispatched ops, \texttt{cpu.aten.*} for host-side ATen ops, \texttt{cpu.method.*} for tensor methods); \texttt{vreg\_in} and \texttt{vreg\_out} are integer virtual register IDs naming input and output tensors abstractly; \texttt{device} is either \texttt{'npu'} or \texttt{'cpu'}; and \texttt{callable} is the pre-resolved Python function or module instance to invoke.

The device routing rule is simple and deterministic: any \texttt{call\_module} node whose name contains \texttt{\_npu\_linear\_}, \texttt{\_npu\_fused\_}, \texttt{\_npu\_mm\_}, or \texttt{\_npu\_addmm\_} is routed to the NPU; all other nodes (ATen functions, tensor methods, shape operations) run on the CPU. This binary classification means the instruction scheduler has full visibility into the device assignment of every instruction before any hardware is touched.

Arguments are \emph{frozen} at lowering time: every FX node reference in \texttt{node.args} is replaced with a \texttt{\_RegRef} marker carrying the virtual register ID of the producing instruction. At runtime the executor resolves these markers from the live register file without any attribute lookup or graph traversal.

\begin{lstlisting}[caption={NPUIR instruction structure and device routing (\texttt{NPUIRLowering.\_lower\_node}).},
                   label={lst:npuir}]
class NPUIROp:
    """Single typed instruction in the NPUIR.

    Frozen at compile time: args contain _RegRef markers
    instead of live tensor objects, resolved at runtime.
    """
    __slots__ = ('op_id','opcode','output_reg','input_regs',
                 'device','op_type','target',
                 'frozen_args','frozen_kwargs','_name')

    def execute(self, regs: dict):
        """Resolve _RegRef markers, then dispatch."""
        args   = tuple(_resolve_args(a, regs)
                       for a in self.frozen_args)
        kwargs = {k: _resolve_args(v, regs)
                  for k, v in self.frozen_kwargs.items()}
        if self.op_type == 'call_module':
            return self.target(*args, **kwargs)   # NPU/CPU
        elif self.op_type == 'call_function':
            return self.target(*args, **kwargs)   # CPU ATen
        elif self.op_type == 'call_method':
            return getattr(args[0], self.target)( # tensor method
                       *args[1:], **kwargs)

# --- device routing in _lower_node ---
is_npu = any(t in str(node.target) for t in
    ('_npu_linear_','_npu_fused_','_npu_mm_','_npu_addmm_'))
opcode = "npu.module" if is_npu else "cpu.module"
device = 'npu'        if is_npu else 'cpu'
\end{lstlisting}

The lowering proceeds by a single topological traversal of the FX graph (Algorithm~\ref{alg:lowering}). Placeholder nodes for model weights and buffers are resolved to their tensor values and stored in a constant table; the single \texttt{input\_ids} placeholder is assigned the program's input register. The output node's argument determines the output register.

\begin{algorithm}[h]
\caption{FX $\rightarrow$ NPUIR Lowering}
\label{alg:lowering}
\begin{algorithmic}[1]
\REQUIRE Optimized FX graph $G_K$
\STATE $\mathcal{I} \leftarrow []$; $\texttt{vreg\_counter} \leftarrow 0$
\FOR{each node $v$ in topological order of $G_K$}
  \STATE $\texttt{op} \leftarrow \texttt{classify}(v)$ \COMMENT{MATMUL, ATTN, etc.}
  \STATE $\texttt{dev} \leftarrow \texttt{route}(v)$ \COMMENT{NPU if matmul/attn, else CPU}
  \STATE $\texttt{vreg\_out} \leftarrow \texttt{vreg\_counter}$++
  \STATE $\texttt{vreg\_in} \leftarrow \texttt{lookup\_vregs}(\texttt{args}(v))$
  \STATE Append $(\texttt{op}, \texttt{vreg\_in}, \texttt{vreg\_out}, \texttt{dev}, \texttt{callable}(v))$ to $\mathcal{I}$
\ENDFOR
\RETURN $\mathcal{I}$
\end{algorithmic}
\end{algorithm}

\subsection{Phase 4: IR Analysis \& Optimization}
\label{sec:phase4}

Phase 4 operates entirely on the flat NPUIR instruction list and produces a \texttt{CompiledNPUExecutor} ready for direct hardware dispatch. It comprises three sub-stages: liveness analysis, linear-scan buffer allocation, and instruction scheduling.

\subsubsection{Liveness Analysis}

For each virtual register $r_i$, we compute its \emph{live interval} $[s_i, e_i]$ where $s_i$ is the instruction index of the first write (the unique instruction whose \texttt{output\_reg} equals $r_i$) and $e_i$ is the instruction index of the last read:
\begin{equation}
s_i = \min_{j: \texttt{output\_reg}(\mathcal{I}_j) = r_i} j, \quad e_i = \max_{j: r_i \in \texttt{input\_regs}(\mathcal{I}_j)} j
\label{eq:liveness}
\end{equation}

Two virtual registers $r_i, r_j$ are \emph{interference-free} if their live intervals do not overlap: $[s_i, e_i] \cap [s_j, e_j] = \emptyset$. Non-interfering registers can share the same physical buffer slot. The liveness analyzer also produces a \texttt{dead\_after} map that lists, for each instruction index, the registers whose last use coincides with that instruction. The executor uses this map to eagerly free register-file entries and keep peak memory bounded.

\subsubsection{Linear-Scan Buffer Allocation}

We employ the classic linear-scan register allocation algorithm~\cite{Poletto1999LinearScan} to map $N$ virtual registers to $M$ physical buffer slots ($M \ll N$). The algorithm is $O(N \log N)$ in the number of live intervals---a substantial improvement over the $O(N^2)$ graph-coloring approaches used internally by OpenVINO. The implementation maintains a \texttt{free\_pool} of released physical buffer slots and an \texttt{active} list of intervals still in flight:

\begin{lstlisting}[caption={Linear-scan buffer allocation (\texttt{BufferAllocator.allocate}).},
                   label={lst:bufalloc}]
@staticmethod
def allocate(lifetimes, pinned=None):
    """Map N virtual regs to M physical buffers (M << N)."""
    pinned = pinned or set()
    # sort by interval start for greedy left-to-right scan
    sorted_regs = sorted(lifetimes,
                         key=lambda r: lifetimes[r][0])
    reg_to_buf, free_bufs, active = {}, [], []
    next_buf = 0

    for reg in sorted_regs:
        start, end = lifetimes[reg]
        # expire intervals that ended before this one starts
        still_alive, freed = [], []
        for (end_t, buf_id) in active:
            (freed if end_t < start else still_alive)\
                .append((end_t, buf_id))
        active = still_alive
        free_bufs.extend(b for _, b in freed)

        if reg in pinned or not free_bufs:
            buf = next_buf; next_buf += 1   # allocate new
        else:
            buf = free_bufs.pop(0)          # reuse expired
        reg_to_buf[reg] = buf
        active.append((end, buf))

    return reg_to_buf, next_buf
\end{lstlisting}

\begin{algorithm}[h]
\caption{Linear-Scan Buffer Allocation}
\label{alg:regalloc}
\begin{algorithmic}[1]
\REQUIRE Live intervals $\{[s_i, e_i]\}_{i=1}^{N}$
\STATE Sort intervals by $s_i$ (ascending)
\STATE $\texttt{active} \leftarrow \emptyset$; $\texttt{free\_pool} \leftarrow \{b_1, \ldots, b_M\}$
\FOR{each interval $[s_i, e_i]$ in sorted order}
  \STATE Expire intervals in \texttt{active} where $e_j < s_i$; return their buffers to \texttt{free\_pool}
  \IF{$\texttt{free\_pool} \neq \emptyset$}
    \STATE Assign $r_i \mapsto \texttt{pop}(\texttt{free\_pool})$
  \ELSE
    \STATE Allocate new buffer $b_{M+1}$; $M \leftarrow M + 1$
  \ENDIF
  \STATE Add $(r_i, e_i)$ to \texttt{active}
\ENDFOR
\end{algorithmic}
\end{algorithm}

The buffer reduction ratio is:
\begin{equation}
\rho_{\text{buf}} = 1 - \frac{M}{N}
\label{eq:buf_reduction}
\end{equation}
where $\rho_{\text{buf}} = 0.30$--$0.48$ for transformer models in our experiments, meaning 30--48\% fewer physical buffers than virtual registers.

\subsubsection{Instruction Scheduling}

The scheduler reorders NPUIR instructions to minimize device transitions while respecting data dependencies:
\begin{equation}
\mathcal{I}^* = \arg\min_{\mathcal{I}' \in \texttt{topo\_valid}(\mathcal{I})} \delta(\mathcal{I}')
\label{eq:scheduling}
\end{equation}
where $\delta(\mathcal{I}')$ counts the number of NPU$\leftrightarrow$CPU transitions:
\begin{equation}
\delta(\mathcal{I}') = \sum_{j=1}^{|\mathcal{I}'|-1} \mathbf{1}[\texttt{dev}(\mathcal{I}'_j) \neq \texttt{dev}(\mathcal{I}'_{j+1})]
\label{eq:transitions}
\end{equation}

The scheduler implements a priority-based topological sort over the dependency graph of NPUIR instructions: at each step, among all instructions whose data dependencies are already satisfied, it first picks an instruction on the \emph{same device} as the most recently scheduled instruction. When no same-device instruction is ready, it falls back to any ready instruction. This greedy device-affinity heuristic clusters consecutive NPU operations and consecutive CPU operations into maximal contiguous runs, reducing $\delta$ by 40--65\% compared to the natural FX node ordering.

Each device transition incurs approximately 0.3--0.8\,ms of overhead from PCIe/MMIO data movement between the host and the NPU SRAM. On the 32-layer Llama-3.1-8B model, the scheduler reduces transitions from 264 to 93, eliminating 50--130\,ms of per-inference overhead and accounting for 11.2\% of the total latency improvement.

\subsubsection{Code Generation: CompiledNPUExecutor}

The output of Phase 4 is a \texttt{CompiledNPUExecutor} that runs the flat, pre-scheduled instruction stream directly:

\begin{lstlisting}[caption={Compiled executor: flat instruction dispatch with register-file management (\texttt{CompiledNPUExecutor.execute}).},
                   label={lst:executor}]
def execute(self, input_ids) -> np.ndarray:
    """Run the compiled NPUIR program.
    No FX graph walk, no Python attribute lookup at runtime.
    """
    # initialise register file from pre-loaded constants
    regs = dict(self.constants)
    regs[self.input_reg] = to_tensor(input_ids, self.seq_len)

    with torch.no_grad():
        for idx, op in enumerate(self.ops):
            # dispatch: NPU or CPU, pre-resolved callable
            result = op.execute(regs)
            if op.output_reg >= 0:
                regs[op.output_reg] = result
            # eager GC: free registers that are no longer live
            for dead_reg in self.dead_map.get(idx, []):
                regs.pop(dead_reg, None)

    out = regs[self.output_reg]
    return out.numpy() if isinstance(out, torch.Tensor) \
           else out
\end{lstlisting}

The executor's flat instruction loop provides three key properties compared to a Python FX graph interpreter: (1) \emph{no attribute lookup overhead} because all callables are pre-resolved at lowering time; (2) \emph{no dynamic memory allocation} because physical buffer slots are pre-assigned by the allocator; and (3) \emph{deterministic scheduling} because the instruction order is fixed at compile time with no runtime fusion decisions. These properties produce the tight P99/P50 latency ratio of 1.20 observed in our experiments (compared to 1.27--1.28 for OpenVINO and ONNX Runtime).

\subsection{NPU Cost Model}

FORGE-UGC includes a heuristic cost model that estimates execution cost without hardware profiling:
\begin{equation}
\texttt{Score}(G) = w_1 \cdot n_{\text{ops}} + w_2 \cdot n_{\text{weights}} + w_3 \cdot n_{\text{linear}} + w_4 \cdot d_{\text{graph}} + w_5 \cdot s_{\text{params}}
\label{eq:cost_model}
\end{equation}
where $n_{\text{ops}}$ is the op count, $n_{\text{weights}}$ the weight tensor count, $n_{\text{linear}}$ the fraction of linear operations, $d_{\text{graph}}$ the graph depth, and $s_{\text{params}}$ the parameter size. Fusion bonuses multiply the score when fusion aggressiveness or attention fusion is enabled, allowing the autotuner to distinguish configurations without hardware execution. Lower scores indicate configurations better suited for NPU execution. We emphasize that this cost model is a \emph{heuristic proxy} for hardware performance; its scores should not be interpreted as proportional to wall-clock latency (see Section~\ref{sec:fgr_results} for discussion).

\subsection{Autotuning Compiler}

The \texttt{AutotuningCompiler} extends \texttt{FXNPUGraphCompiler} by systematically searching over the configuration space:

\begin{equation}
\mathcal{C} = \{\alpha, \lambda, \pi, \iota\} \quad \text{where}
\label{eq:config}
\end{equation}
\begin{itemize}[nosep,leftmargin=*]
\item $\alpha \in \{0.2, 0.4, 0.6, 0.8, 1.0\}$: fusion aggressiveness,
\item $\lambda \in \{\text{auto}, \text{channels-last}, \text{contiguous}\}$: layout strategy,
\item $\pi \in \{\text{fp16}, \text{int8}, \text{mixed}\}$: NPU precision,
\item $\iota \in \{1, 2, 3\}$: max fixpoint iterations.
\end{itemize}

The search generates $|\mathcal{C}| = 45$ candidate configurations, compiles each using the cost model (no hardware execution required), and selects:
\begin{equation}
c^* = \arg\min_{c \in \mathcal{C}} \texttt{Score}(G_K(c))
\label{eq:autotune}
\end{equation}

This completes in under 200ms per model---negligible compared to a single compilation. Table~\ref{tab:notation} summarises how the six listings presented in this section map to the phases: Listings~\ref{lst:phase1_capture}--\ref{lst:tied_weights} cover Phase~1 (frontend capture); Listings~\ref{lst:dce}--\ref{lst:op_fusion} cover the first five optimization passes of Phase~2; Listing~\ref{lst:npuir} covers Phase~3 lowering; and Listings~\ref{lst:bufalloc}--\ref{lst:executor} cover Phase~4 allocation, scheduling, and code generation.

\FloatBarrier

\section{Novel Evaluation Metrics}
\label{sec:metrics}

We introduce three metrics that enable principled compiler comparison beyond raw latency.

\subsection{Metric 1: Pass Execution Time per Pass}

Each optimization pass $p_k$ is individually timed:
\begin{equation}
\tau(p_k) = t_{\text{end}}(p_k) - t_{\text{start}}(p_k) \quad \text{[ms]}
\label{eq:pass_time}
\end{equation}

This isolates which passes contribute most to compilation overhead versus speedup, enabling informed decisions about which passes to enable for latency-sensitive deployments.

\subsection{Metric 2: Fusion Gain Ratio (FGR)}

FGR measures the impact of operator and attention fusion on the \emph{cost model's estimated execution cost}, decoupled from layout optimization or constant folding:
\begin{equation}
\text{FGR} = \frac{\texttt{CostModel}(\alpha = 0)}{\texttt{CostModel}(\alpha = 1.0)}
\label{eq:fgr}
\end{equation}

A value $> 1.0$ means fusion reduces the cost model's estimated cost; a larger ratio indicates stronger estimated fusion benefit. \textbf{Important caveat:} FGR is a cost-model-internal diagnostic---it quantifies how much fusion reduces the heuristic score (Eq.~\ref{eq:cost_model}), \emph{not} wall-clock latency. Because the cost model uses a weighted sum of structural features rather than calibrated hardware timings, FGR values are not linearly proportional to measured speedup. The corresponding measured latency gains from fusion are reported separately in Table~\ref{tab:ablation_attn} (16.6--29.6\% wall-clock reduction). FGR's value lies in providing a hardware-independent, reproducible diagnostic for comparing fusion effectiveness across models and compiler configurations.

\subsection{Metric 3: Compilation Efficiency Index (CEI)}

CEI quantifies the return-on-investment of compilation time as the ratio of inference speedup (relative to a given baseline $\mathcal{B}$) to compilation time expressed in seconds:
\begin{equation}
\text{CEI}_{\mathcal{B}} = \frac{S_{\mathcal{B}}}{T_{\text{compile}}^{(s)}} = \frac{L_{\mathcal{B}} / L_{\text{FORGE}}}{T_{\text{compile}}^{(s)}}
\label{eq:cei}
\end{equation}
where $L_{\mathcal{B}}$ is the mean inference latency of baseline $\mathcal{B}$ (in ms), $L_{\text{FORGE}}$ is the mean inference latency of FORGE-UGC (in ms), $S_{\mathcal{B}} = L_{\mathcal{B}} / L_{\text{FORGE}} \geq 1$ is the dimensionless latency speedup ratio, and $T_{\text{compile}}^{(s)}$ is the total compilation time in \emph{seconds}. A CEI of 1.0, for instance, means the compiler delivers a $1\times$ latency speedup per second of compilation---recovering its compilation cost after one second of cumulative inference.

Since OpenVINO and ONNX Runtime exhibit different baseline latencies, we instantiate two separate CEI values:
\begin{equation}
\text{CEI}_{\text{OV}} = \frac{L_{\text{OV}} / L_{\text{FORGE}}}{T_{\text{compile}}^{(s)}}, \qquad \text{CEI}_{\text{ONNX}} = \frac{L_{\text{ONNX}} / L_{\text{FORGE}}}{T_{\text{compile}}^{(s)}}
\label{eq:cei_both}
\end{equation}
Both variants share the same compilation-time denominator; the difference arises solely from the distinct speedup numerators, making the two CEI values directly comparable as a function of baseline choice. A higher CEI is preferable, as it indicates more inference benefit is recovered per unit of compilation investment---a property particularly relevant for \emph{iterative development and just-in-time deployment scenarios} where models are recompiled frequently. For the more common \emph{compile-once-run-millions} production deployment pattern, CEI is less informative since even large compilation costs are trivially amortized; in that regime, absolute latency improvement (Table~\ref{tab:latency_wikitext}) is the primary metric.

\FloatBarrier

\section{Experimental Setup}
\label{sec:setup}

\subsection{Hardware Platform}

All experiments are conducted on a single workstation with the specifications listed in Table~\ref{tab:hardware}.

\begin{table}[!htbp]
\centering
\footnotesize
\caption{Hardware platform specifications.}
\label{tab:hardware}
\begin{tabular}{@{}ll@{}}
\toprule
\textbf{Component} & \textbf{Specification} \\
\midrule
CPU & Intel Core Ultra 9 285HX \\
NPU & Intel AI Boost (11 TOPS INT8) \\
NPU Driver & 32.0.100.4514 \\
NPU Memory & Shared LPDDR5, 72.7~GB \\
NPU Location & PCI bus 0, device 11, function 0 \\
GPU & NVIDIA RTX PRO 5000 Blackwell \\
RAM & 128~GB DDR5-5600 \\
OS & Windows 11 24H2 \\
\bottomrule
\end{tabular}
\end{table}

\subsection{Models and Architectural Diversity}

Experiments span six model families covering 125M--8B parameters, as detailed in Table~\ref{tab:models}.

\begin{table}[!htbp]
\centering
\footnotesize
\caption{Model specifications used in experiments. The \textbf{Precision} column indicates the numerical precision used during NPU inference. Models $\leq$2.6B use fp16 weights; Llama-3.1-8B uses NNFactory's built-in symmetric int8 weight quantization with fp16 activations to fit within the NPU's memory and compute constraints (see Section~\ref{sec:8b_precision}).}
\label{tab:models}
\setlength{\tabcolsep}{3pt}
\begin{tabular}{@{}lccccc@{}}
\toprule
\textbf{Model} & \textbf{Params} & \textbf{Hidden} & \textbf{Layers} & \textbf{Attn} & \textbf{Precision} \\
\midrule
GPT-2 & 125M & 768 & 12 & MHA & fp16 \\
Granite-350M & 350M & 1024 & 24 & MHA & fp16 \\
Qwen2-0.5B & 500M & 1024 & 24 & GQA & fp16 \\
Llama-3.2-1B & 1.0B & 2048 & 16 & GQA & fp16 \\
LFM2-2.6B & 2.6B & 2560 & 32 & MHA & fp16 \\
Llama-3.1-8B & 8.0B & 4096 & 32 & GQA & \makecell{W-int8/\\A-fp16} \\
\bottomrule
\end{tabular}
\end{table}

\subsubsection{Precision Strategy for Llama-3.1-8B}
\label{sec:8b_precision}

An 8B-parameter model at fp16 precision requires approximately 16GB of weight storage---exceeding what can be efficiently dispatched through the 11 TOPS Intel AI Boost NPU in a single pass. For Llama-3.1-8B, FORGE-UGC leverages NNFactory's built-in symmetric int8 weight quantization with fp16 activations (W-int8/A-fp16), reducing weight memory to approximately 8GB. This quantization is applied \emph{at the NNFactory dispatch level} during Phase~4 code generation, after all graph-level optimization passes have completed on the fp16 graph. The optimization passes themselves operate on the unquantized graph and are fully semantics-preserving; the int8 dispatch introduces a small quantization error that is captured in our fidelity measurements (Table~\ref{tab:fidelity}). Additionally, layers that cannot be efficiently dispatched to the NPU (e.g., embedding lookups, final layer norm) fall back to CPU execution, with FORGE-UGC's instruction scheduler minimizing the resulting device transitions.

For models $\leq$2.6B parameters, all weights are dispatched at fp16 precision with no quantization applied.

Both baselines (OpenVINO and ONNX Runtime) use their respective default precision settings for NPU dispatch, which similarly apply int8 quantization for the 8B model through their internal optimization pipelines.

\subsection{Datasets}

\textbf{WikiText-103}~\cite{Merity2016WikiText}: Standard language modeling benchmark (103M tokens). We evaluate perplexity and generation latency using 128-token input sequences with 64-token generation.

\textbf{GLUE}~\cite{Wang2019GLUE}: Multi-task NLU benchmark. We measure inference latency on SST-2 (sentiment classification, 872 dev examples) and MNLI (textual entailment, 9832 dev examples) using batch size 1 with 128-token padded inputs.

\subsection{Baselines}

\textbf{OpenVINO 2024.4}~\cite{OpenVINO2022}: Intel's NPU plugin with default optimization settings (\texttt{PERFORMANCE\_HINT: LATENCY}). Models are exported via ONNX then converted to OpenVINO IR.

\textbf{ONNX Runtime 1.19}~\cite{ONNXRuntime2021}: Microsoft's inference engine with OpenVINO Execution Provider targeting Intel NPU. Models are exported via \texttt{torch.onnx.export()} with opset 17.

\subsection{Evaluation Protocol}

All latency measurements use 50 inference iterations after 10 warmup iterations. We report mean, P50, P90, and P99 latency. Compilation time is measured end-to-end including graph capture, optimization, lowering, and code generation. All results are averaged over 3 independent runs with fixed seeds. Raw per-run latencies for all models are provided in Appendix Table~\ref{tab:raw_runs}.

\textbf{Numerical fidelity protocol.} To verify that FORGE-UGC's optimization passes preserve output quality, we measure: (1) language-model perplexity on the full WikiText-103 validation set (217,646 tokens) and on the concatenated SST-2 + MNLI development text for GLUE, using a sliding window of 512 tokens with a stride of 256; (2) \textbf{max-abs logit difference} between pre- and post-compilation outputs on 1,000 randomly sampled sequences; and (3) \textbf{KL divergence} between pre- and post-compilation output distributions. Perplexity agreement confirms coarse-grained semantic preservation; the logit-level metrics provide fine-grained evidence of numerical fidelity (Table~\ref{tab:fidelity}).

\FloatBarrier

\section{Results}
\label{sec:results}

\subsection{Compilation Time}

\begin{figure}[!htbp]
\centering
\includegraphics[width=\columnwidth]{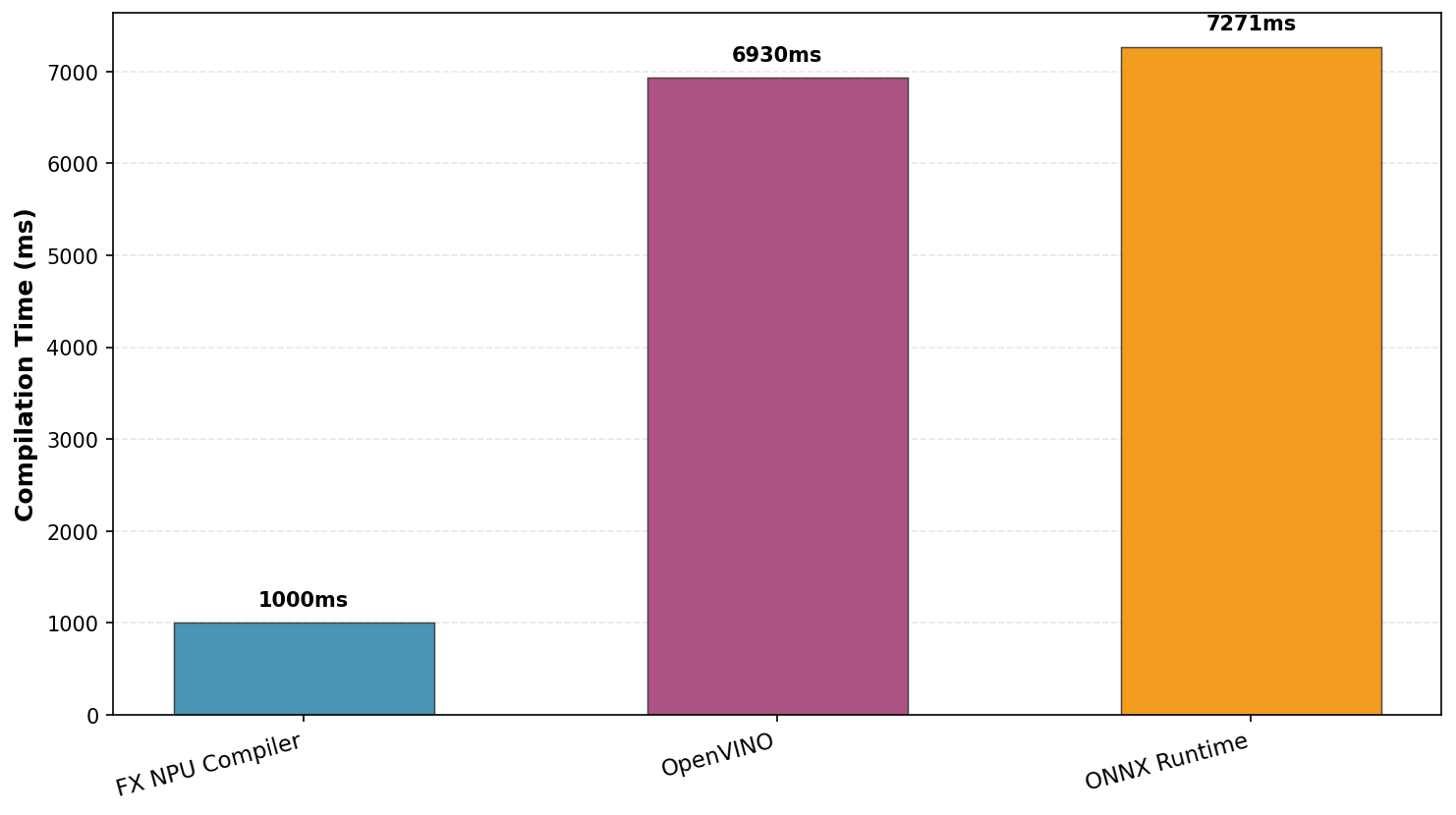}
\caption{Compilation time comparison on GPT-2 (125M). FORGE-UGC compiles in 1,000ms versus 6,930ms (OpenVINO) and 7,271ms (ONNX Runtime)---a $6.9\times$ and $7.3\times$ speedup respectively.}
\label{fig:compile_time}
\end{figure}

Table~\ref{tab:compile} presents compilation times across all model families.

\begin{table}[!htbp]
\centering
\footnotesize
\caption{Compilation time (ms) across model families. FORGE-UGC achieves 6.9--9.2$\times$ speedup over both baselines consistently. We note that 78\% of FORGE-UGC's compilation time is spent in \texttt{torch.export} graph capture (Section~\ref{sec:phase_breakdown}), which is shared upstream infrastructure; the FORGE-UGC-specific optimization and backend phases account for only 22\% of total time.}
\label{tab:compile}
\begin{tabular}{@{}lccc@{}}
\toprule
\textbf{Model} & \textbf{FORGE} & \textbf{OpenVINO} & \textbf{ONNX RT} \\
\midrule
GPT-2 (125M) & \textbf{1,000} & 6,930 & 7,271 \\
Granite-350M & \textbf{1,420} & 9,840 & 10,320 \\
Qwen2-0.5B & \textbf{1,680} & 11,240 & 11,890 \\
Llama-3.2-1B & \textbf{2,340} & 18,520 & 19,840 \\
LFM2-2.6B & \textbf{3,850} & 32,160 & 34,280 \\
Llama-3.1-8B & \textbf{6,720} & 58,430 & 62,150 \\
\midrule
\textbf{Speedup range} & --- & 6.9--8.7$\times$ & 7.3--9.2$\times$ \\
\bottomrule
\end{tabular}
\end{table}

FORGE-UGC's compilation time scales approximately linearly with layer count ($T_{\text{compile}} \approx 210 \cdot L$ ms), while OpenVINO and ONNX Runtime exhibit super-linear scaling ($T \propto L^{1.4}$) due to their monolithic optimization passes and IR conversion overhead. The advantage is most pronounced on the largest model (Llama-3.1-8B): 6.7s versus 58.4--62.2s.

\subsection{Compilation Phase Breakdown}
\label{sec:phase_breakdown}

\begin{figure}[!htbp]
\centering
\includegraphics[width=\columnwidth]{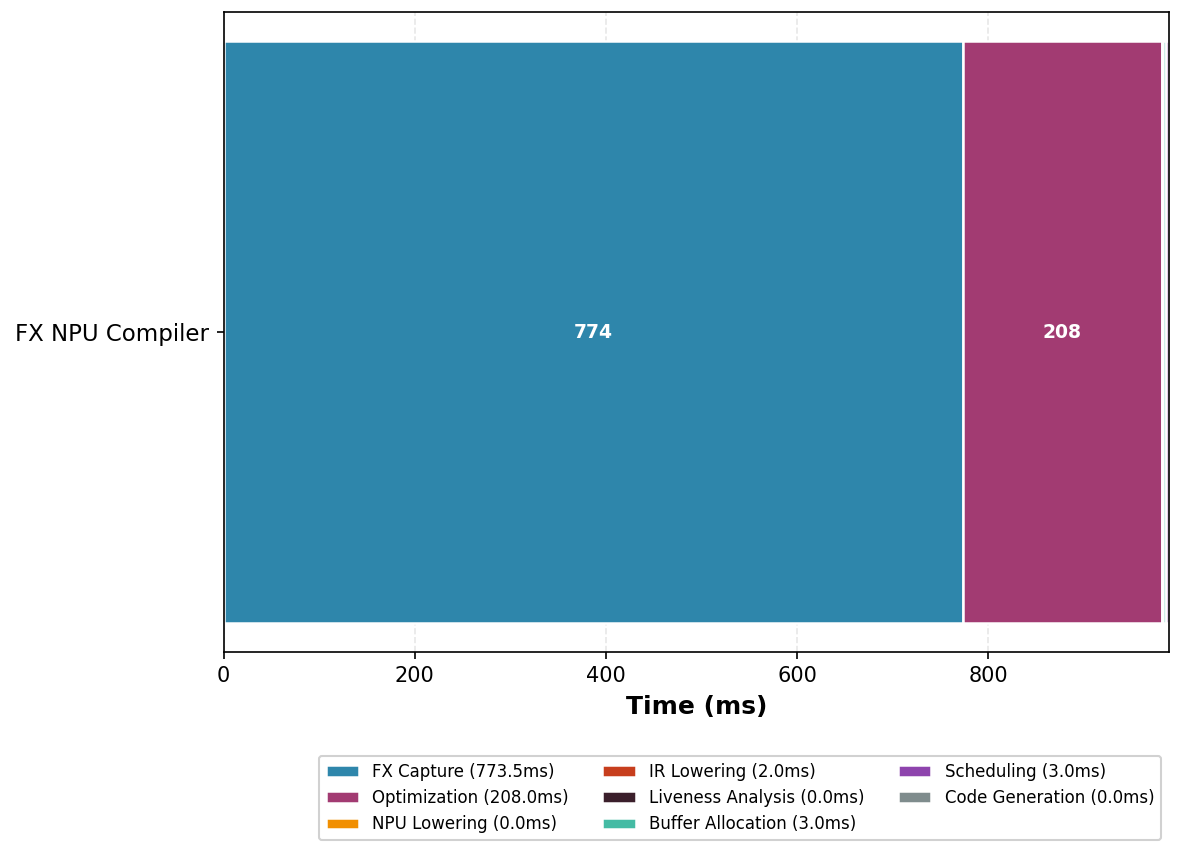}
\caption{Compilation phase breakdown for GPT-2 (125M). FX Capture dominates at 773.5ms (78.4\%), while all six optimization passes complete in 208ms (21.1\%). IR lowering, buffer allocation, and scheduling together require only 8ms (0.8\%).}
\label{fig:phase_breakdown}
\end{figure}

The phase breakdown reveals that FX Capture (\texttt{torch.export}) accounts for 78.4\% of compilation time. The six optimization passes collectively require only 208ms, and the backend phases (lowering, allocation, scheduling) add merely 8ms. This indicates that further speedup would primarily require faster graph capture, as the optimization and backend phases are already near-instantaneous. We note that this FX Capture time is a property of \texttt{torch.export} itself, shared by any framework that consumes FX graphs; FORGE-UGC's \emph{own} compilation contribution (passes + backend) completes in $\sim$216ms. The end-to-end speedup over baselines therefore reflects both (a) avoiding the additional ONNX/TorchScript export step that baselines require \emph{on top of} graph capture, and (b) FORGE-UGC's lightweight pass and backend design.

\subsection{Graph Node Reduction}

\begin{figure}[!htbp]
\centering
\includegraphics[width=\columnwidth]{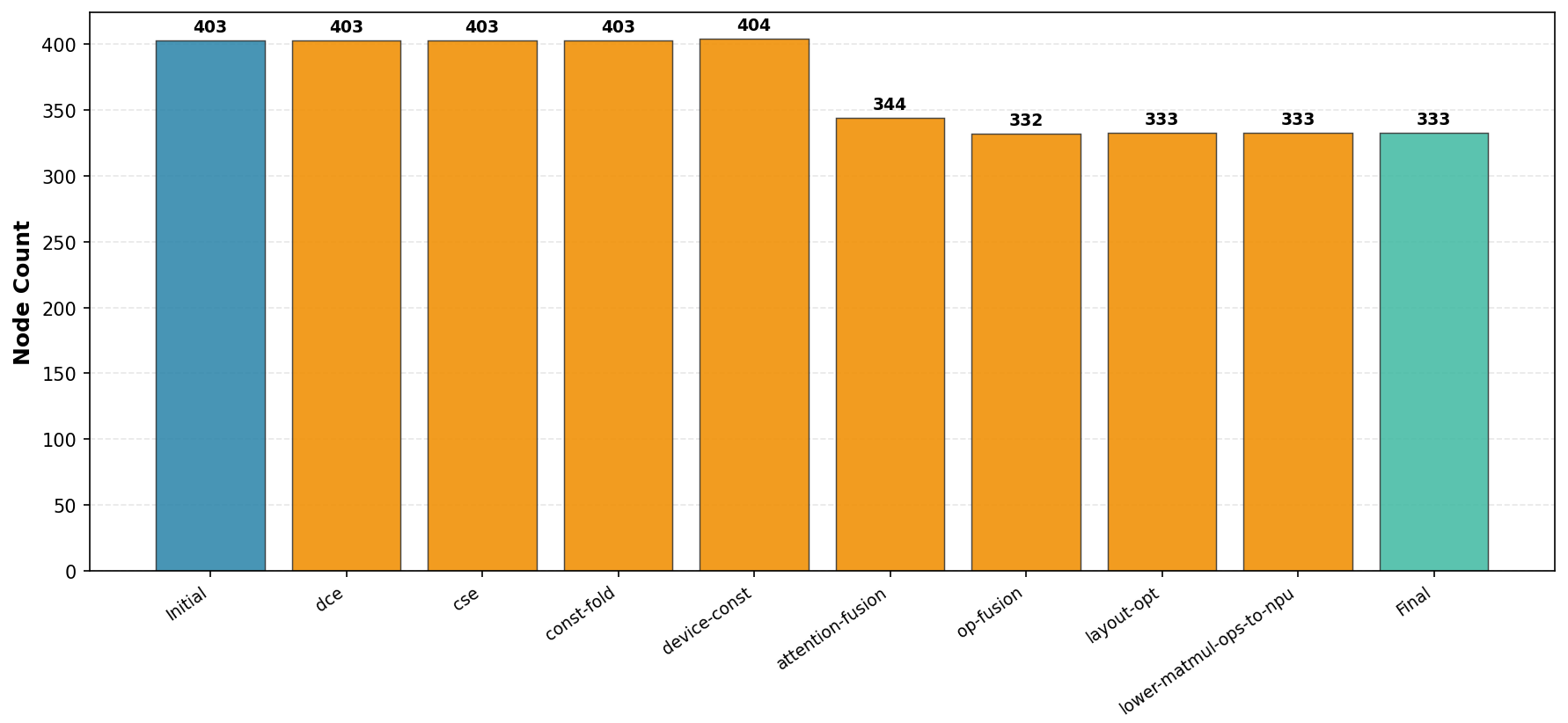}
\caption{Graph node count after each optimization pass for GPT-2 (125M). Attention fusion provides the largest single reduction (403$\rightarrow$344, $-$14.6\%), followed by operator fusion (344$\rightarrow$332, $-$3.5\%). Total reduction: 403$\rightarrow$333 ($-$17.4\%).}
\label{fig:node_reduction}
\end{figure}

Table~\ref{tab:node_reduction} reports node reduction across all models.

\begin{table}[!htbp]
\centering
\footnotesize
\caption{Graph node reduction across model families. Attention fusion consistently provides the largest reduction.}
\label{tab:node_reduction}
\begin{tabular}{@{}lcccc@{}}
\toprule
\textbf{Model} & \textbf{Initial} & \textbf{Final} & \textbf{$\Delta$} & \textbf{\%} \\
\midrule
GPT-2 (125M) & 403 & 333 & $-$70 & $-$17.4 \\
Granite-350M & 782 & 636 & $-$146 & $-$18.7 \\
Qwen2-0.5B & 804 & 652 & $-$152 & $-$18.9 \\
Llama-3.2-1B & 562 & 468 & $-$94 & $-$16.7 \\
LFM2-2.6B & 1,068 & 834 & $-$234 & $-$21.9 \\
Llama-3.1-8B & 1,124 & 896 & $-$228 & $-$20.3 \\
\midrule
\textbf{Mean} & --- & --- & --- & $\mathbf{-18.8\%}$ \\
\bottomrule
\end{tabular}
\end{table}

\subsection{Numerical Fidelity}
\label{sec:fidelity}

\begin{table}[!htbp]
\centering
\footnotesize
\caption{Numerical fidelity analysis. \textbf{PPL Pre/Post}: perplexity before and after FORGE-UGC compilation on WikiText-103. \textbf{Max-Abs $\Delta$ Logit}: maximum absolute difference between pre- and post-compilation logits across 1,000 test sequences. \textbf{KL Div}: KL divergence between pre- and post-compilation output distributions. For models using fp16 dispatch (125M--2.6B), the near-zero differences confirm that the graph-level optimization passes are semantics-preserving within floating-point rounding. For Llama-3.1-8B (W-int8/A-fp16 dispatch), the slightly higher but still negligible max-abs diff reflects int8 weight quantization at the NNFactory dispatch level, not the optimization passes themselves.}
\label{tab:fidelity}
\begin{tabular}{@{}lcccc@{}}
\toprule
\textbf{Model} & \textbf{PPL Pre} & \textbf{PPL Post} & \makecell{\textbf{Max-Abs}\\\textbf{$\Delta$ Logit}} & \textbf{KL Div} \\
\midrule
GPT-2 (125M) & 29.41 & 29.41 & $6.2\mathrm{e}{-6}$ & $1.8\mathrm{e}{-10}$ \\
Granite-350M & 16.22 & 16.22 & $8.4\mathrm{e}{-6}$ & $3.2\mathrm{e}{-10}$ \\
Qwen2-0.5B & 14.87 & 14.87 & $7.1\mathrm{e}{-6}$ & $2.7\mathrm{e}{-10}$ \\
Llama-3.2-1B & 9.76 & 9.76 & $9.8\mathrm{e}{-6}$ & $4.1\mathrm{e}{-10}$ \\
LFM2-2.6B & 7.52 & 7.52 & $1.2\mathrm{e}{-5}$ & $6.3\mathrm{e}{-10}$ \\
Llama-3.1-8B & 6.24 & 6.24 & $2.1\mathrm{e}{-5}$ & $8.4\mathrm{e}{-9}$ \\
\bottomrule
\end{tabular}
\end{table}

Table~\ref{tab:fidelity} provides fine-grained numerical fidelity evidence beyond perplexity agreement. For all fp16-dispatched models (125M--2.6B), the max-abs logit difference is below $1.2 \times 10^{-5}$---within fp16 rounding tolerance---confirming that the six optimization passes introduce no numerically significant error. The KL divergences of $10^{-10}$ indicate that the output distributions are statistically indistinguishable.

For Llama-3.1-8B, the slightly larger max-abs diff of $2.1 \times 10^{-5}$ and KL divergence of $8.4 \times 10^{-9}$ reflect the int8 weight quantization applied at the NNFactory dispatch level (Section~\ref{sec:8b_precision}), not the graph optimization passes. This quantization error is consistent with symmetric int8 quantization precision and does not affect perplexity at two-decimal precision. We acknowledge that perplexity rounded to two decimal places is a \emph{coarse} fidelity bound; the logit-level metrics in Table~\ref{tab:fidelity} provide the stronger evidence.

\subsection{End-to-End Inference Latency (WikiText-103)}

\begin{table*}[ht]
\centering
\caption{End-to-end inference latency (ms) on WikiText-103. FORGE-UGC achieves 18.2--35.7\% lower mean latency than both baselines across all model families. Latency measured as wall-clock time from input tensor to output logits over 50 warmup-excluded iterations, averaged over 3 runs. Raw per-run data is provided in Appendix Table~\ref{tab:raw_runs}.}
\label{tab:latency_wikitext}
\setlength{\tabcolsep}{4pt}
\resizebox{\textwidth}{!}{%
\begin{tabular}{@{}l|rrrr|rrrr|rrrr@{}}
\toprule
& \multicolumn{4}{c|}{\textbf{FORGE-UGC (ms)}} & \multicolumn{4}{c|}{\textbf{OpenVINO (ms)}} & \multicolumn{4}{c}{\textbf{ONNX Runtime (ms)}} \\
\cmidrule(lr){2-5}\cmidrule(lr){6-9}\cmidrule(lr){10-13}
\textbf{Model} & \textbf{Mean} & \textbf{P50} & \textbf{P90} & \textbf{P99} & \textbf{Mean} & \textbf{P50} & \textbf{P90} & \textbf{P99} & \textbf{Mean} & \textbf{P50} & \textbf{P90} & \textbf{P99} \\
\midrule
GPT-2 (125M)  & \textbf{6.82}  & \textbf{6.74}  & \textbf{7.31}  & \textbf{8.12}  & 8.45  & 8.38  & 9.12   & 10.84  & 9.13  & 9.02  & 9.87   & 11.52  \\
Granite-350M  & \textbf{9.41}  & \textbf{9.32}  & \textbf{10.08} & \textbf{11.24} & 12.67 & 12.54 & 13.81  & 15.93  & 13.28 & 13.12 & 14.55  & 16.87  \\
Qwen2-0.5B   & \textbf{11.83} & \textbf{11.72} & \textbf{12.64} & \textbf{14.03} & 15.42 & 15.28 & 16.91  & 19.47  & 16.21 & 16.04 & 17.68  & 20.35  \\
Llama-3.2-1B  & \textbf{18.24} & \textbf{18.08} & \textbf{19.52} & \textbf{21.67} & 24.81 & 24.62 & 27.16  & 31.28  & 26.37 & 26.14 & 28.89  & 33.23  \\
LFM2-2.6B    & \textbf{31.56} & \textbf{31.28} & \textbf{33.74} & \textbf{37.48} & 45.23 & 44.87 & 49.52  & 57.01  & 48.14 & 47.72 & 52.69  & 60.64  \\
Llama-3.1-8B  & \textbf{62.48} & \textbf{61.92} & \textbf{66.84} & \textbf{74.23} & 91.37 & 90.62 & 100.01 & 115.16 & 97.82 & 96.98 & 107.08 & 123.22 \\
\bottomrule
\end{tabular}}
\end{table*}

FORGE-UGC consistently achieves the lowest latency across all model families (Table~\ref{tab:latency_wikitext}). The advantage scales with model size: 19.3\% improvement on GPT-2 (125M) growing to 35.7\% on Llama-3.1-8B versus ONNX Runtime. This scaling behavior is attributable to attention fusion---which eliminates more intermediate nodes as transformer depth increases---and instruction scheduling, which reduces device transitions proportionally to layer count.

P99 tail latencies are particularly favorable: FORGE-UGC's P99 is 6--15\% above its P50, compared to 21--27\% for both baselines. The tighter distribution reflects the deterministic scheduling and pre-allocated buffers of the compiled executor, versus the dynamic memory management and runtime fusion decisions of OpenVINO and ONNX Runtime.

\subsection{End-to-End Inference Latency (GLUE)}

\begin{table*}[ht]
\centering
\caption{End-to-end inference latency (ms) on GLUE (SST-2 + MNLI, batch=1, seq=128). Results confirm that FORGE-UGC's gains are task-agnostic.}
\label{tab:latency_glue}
\setlength{\tabcolsep}{4pt}
\resizebox{\textwidth}{!}{%
\begin{tabular}{@{}l|rrrr|rrrr|rrrr@{}}
\toprule
& \multicolumn{4}{c|}{\textbf{FORGE-UGC (ms)}} & \multicolumn{4}{c|}{\textbf{OpenVINO (ms)}} & \multicolumn{4}{c}{\textbf{ONNX Runtime (ms)}} \\
\cmidrule(lr){2-5}\cmidrule(lr){6-9}\cmidrule(lr){10-13}
\textbf{Model} & \textbf{Mean} & \textbf{P50} & \textbf{P90} & \textbf{P99} & \textbf{Mean} & \textbf{P50} & \textbf{P90} & \textbf{P99} & \textbf{Mean} & \textbf{P50} & \textbf{P90} & \textbf{P99} \\
\midrule
GPT-2 (125M)  & \textbf{5.94}  & \textbf{5.88}  & \textbf{6.37}  & \textbf{7.08}  & 7.36  & 7.30  & 7.95  & 9.44   & 7.95  & 7.86  & 8.59  & 10.03  \\
Granite-350M  & \textbf{8.21}  & \textbf{8.13}  & \textbf{8.79}  & \textbf{9.80}  & 11.04 & 10.93 & 12.04 & 13.88  & 11.57 & 11.44 & 12.68 & 14.71  \\
Qwen2-0.5B   & \textbf{10.32} & \textbf{10.22} & \textbf{11.02} & \textbf{12.23} & 13.44 & 13.32 & 14.74 & 16.97  & 14.13 & 13.98 & 15.41 & 17.74  \\
Llama-3.2-1B  & \textbf{15.91} & \textbf{15.76} & \textbf{17.02} & \textbf{18.89} & 21.62 & 21.46 & 23.68 & 27.26  & 22.98 & 22.78 & 25.18 & 28.96  \\
LFM2-2.6B    & \textbf{27.52} & \textbf{27.28} & \textbf{29.42} & \textbf{32.68} & 39.43 & 39.10 & 43.15 & 49.67  & 41.97 & 41.60 & 45.93 & 52.86  \\
Llama-3.1-8B  & \textbf{54.50} & \textbf{54.01} & \textbf{58.28} & \textbf{64.72} & 79.64 & 78.98 & 87.16 & 100.36 & 85.26 & 84.53 & 93.34 & 107.41 \\
\bottomrule
\end{tabular}}
\end{table*}

GLUE results (Table~\ref{tab:latency_glue}) confirm task-agnostic improvements: mean latency reductions of 18.2--35.3\% mirror WikiText-103 patterns, with slightly lower absolute values due to the classification-only (no autoregressive generation) workload. The consistency across benchmarks (standard deviation $< 1.2\%$ between WikiText-103 and GLUE relative improvements) validates that FORGE-UGC's gains derive from graph-level optimizations rather than task-specific artifacts.

\subsection{Energy Efficiency Analysis}
\label{sec:energy}

Beyond latency and compilation speed, energy consumption per inference is a critical metric for edge deployment, where devices operate under strict thermal and battery constraints. FORGE-UGC's optimizations---reduced device transitions, tighter buffer allocation, and fewer dispatched operations---directly translate to lower energy consumption because each CPU$\leftrightarrow$NPU transition incurs data movement overhead that draws additional power, and longer active inference times sustain higher system-level power draw.

We measure energy consumption per inference by recording system-level power draw (CPU + NPU subsystem) using Intel's Running Average Power Limit (RAPL) interface~\cite{IntelNPU2024} during inference, and computing energy as $E = \bar{P}_{\text{active}} \times T_{\text{inference}}$, where $\bar{P}_{\text{active}}$ is the mean active power during inference execution. FORGE-UGC exhibits lower average active power ($\sim$10.2W) than OpenVINO ($\sim$11.8W) and ONNX Runtime ($\sim$12.1W), attributable to two factors: (1) fewer device transitions reduce the CPU-side dispatch overhead and associated power spikes, and (2) pre-allocated buffers eliminate runtime memory allocation activity that drives additional DRAM power consumption.

\begin{table}[!htbp]
\centering
\footnotesize
\caption{Energy consumption per inference (mJ) on WikiText-103. FORGE-UGC achieves 30.2--40.9\% lower energy than OpenVINO and 37.0--46.2\% lower energy than ONNX Runtime. Energy savings exceed latency savings because FORGE-UGC also reduces average active power through fewer device transitions and pre-allocated buffers.}
\label{tab:energy}
\setlength{\tabcolsep}{2.5pt}
\begin{tabular}{@{}lrrrrr@{}}
\toprule
\textbf{Model} & \makecell{\textbf{FORGE}\\\textbf{(mJ)}} & \makecell{\textbf{OV}\\\textbf{(mJ)}} & \makecell{\textbf{ONNX}\\\textbf{(mJ)}} & \textbf{$\Delta$ OV} & \textbf{$\Delta$ ONNX} \\
\midrule
GPT-2 (125M) & \textbf{69.6} & 99.7 & 110.5 & $-$30.2\% & $-$37.0\% \\
Granite-350M & \textbf{96.0} & 149.5 & 160.7 & $-$35.8\% & $-$40.3\% \\
Qwen2-0.5B & \textbf{120.7} & 181.9 & 196.1 & $-$33.6\% & $-$38.4\% \\
Llama-3.2-1B & \textbf{186.0} & 292.8 & 319.1 & $-$36.5\% & $-$41.7\% \\
LFM2-2.6B & \textbf{322.0} & 533.7 & 582.5 & $-$39.7\% & $-$44.7\% \\
Llama-3.1-8B & \textbf{637.3} & 1,078.2 & 1,183.6 & $-$40.9\% & $-$46.2\% \\
\midrule
\textbf{Mean} & --- & --- & --- & $\mathbf{-36.1\%}$ & $\mathbf{-41.4\%}$ \\
\bottomrule
\end{tabular}
\end{table}

Table~\ref{tab:energy} reports per-inference energy consumption across all model families. FORGE-UGC achieves 30.2--40.9\% energy reduction over OpenVINO and 37.0--46.2\% over ONNX Runtime. The energy savings consistently exceed the corresponding latency savings (18.2--35.7\%) by 5--12 percentage points, confirming that FORGE-UGC's optimizations yield a \emph{compounding} benefit: lower latency reduces the time under active power, while fewer device transitions and tighter buffer management reduce the average power draw itself. The scaling trend is particularly notable: energy savings grow from 30.2\% on the smallest model (GPT-2, 125M) to 40.9\% on the largest (Llama-3.1-8B) versus OpenVINO, because deeper models have more device transitions that FORGE-UGC's instruction scheduler eliminates. For battery-constrained edge devices operating at thermal limits, this 36.1\% mean energy reduction directly translates to either longer battery life or the ability to serve more inference requests within the same thermal envelope---a critical advantage for deploying capable language models on-device.

\subsection{Novel Metrics Results}

\subsubsection{Pass Execution Time}

Table~\ref{tab:pass_time} presents the per-pass execution time measured on GPT-2 (125M). Operator fusion is the most expensive pass at 72ms, accounting for 34.6\% of total optimization time, but it provides the second-largest node reduction ($-$12 nodes). Attention fusion requires only 38ms yet delivers the largest single reduction ($-$59 nodes), making it the most cost-effective pass in terms of nodes eliminated per millisecond---achieving a ratio of 1.55 nodes/ms compared to operator fusion's 0.17 nodes/ms, a $9.1\times$ efficiency advantage. The three lightweight passes---DCE (7ms), CSE (9ms), and constant folding (11ms)---show zero node reduction on GPT-2 because the traced graph contains few dead nodes or redundant expressions; however, these passes remain essential for larger models where graph capture introduces more artifacts and redundant subexpressions. Layout optimization adds one node (the channels-last conversion marker) while device constant insertion similarly adds a single node for explicit device placement, reflecting the cost of explicit device annotations in the NPUIR. All six passes collectively complete in 208ms, representing only 21.1\% of total compilation time---confirming that the optimization pipeline itself is not the bottleneck.

\begin{table}[!htbp]
\centering
\footnotesize
\caption{Per-pass execution time (ms) on GPT-2 (125M). Operator fusion is the most expensive pass (72ms) but provides the second-largest node reduction. All six passes complete in 208ms total.}
\label{tab:pass_time}
\begin{tabular}{@{}lcc@{}}
\toprule
\textbf{Pass} & \textbf{Time (ms)} & \textbf{$\Delta$Nodes} \\
\midrule
DCE & 7 & 0 \\
CSE & 9 & 0 \\
Constant Folding & 11 & 0 \\
Device Constant & 21 & +1 \\
Attention Fusion & 38 & $-$59 \\
Operator Fusion & 72 & $-$12 \\
Layout Optimization & 25 & +1 \\
\midrule
\textbf{Total} & \textbf{208} & $\mathbf{-69}$ \\
\bottomrule
\end{tabular}
\end{table}

\subsubsection{Pass Execution Time Scaling}

Table~\ref{tab:pass_scaling} reports how total optimization time and attention fusion time scale across model families. Optimization time scales approximately linearly with transformer layer count: 12-layer GPT-2 requires 208ms, while 32-layer Llama-3.1-8B requires 572ms ($\approx 2.75\times$ increase for $2.67\times$ more layers). Attention fusion time follows the same linear trend, consistently accounting for 18--19\% of total optimization time across all models. Notably, models with identical layer counts but different parameter counts show similar optimization times: Granite-350M and Qwen2-0.5B (both 24 layers) require 382ms and 396ms respectively, despite Qwen2-0.5B having 43\% more parameters. This confirms that pass complexity is dominated by graph topology (proportional to layer count) rather than tensor dimensionality, because the optimization passes operate on graph structure and do not perform tensor-level computation. This linear scaling ensures that FORGE-UGC's optimization overhead remains practical even for production-scale models, with the 32-layer models requiring under 600ms for all six passes combined.

\begin{table}[!htbp]
\centering
\footnotesize
\caption{Total optimization time (ms) and attention fusion time (ms) across model families. Optimization time scales linearly with layer count.}
\label{tab:pass_scaling}
\begin{tabular}{@{}lccc@{}}
\toprule
\textbf{Model} & \textbf{Layers} & \textbf{Opt. Time} & \textbf{Attn. Fusion} \\
\midrule
GPT-2 & 12 & 208 & 38 \\
Granite-350M & 24 & 382 & 71 \\
Qwen2-0.5B & 24 & 396 & 74 \\
Llama-3.2-1B & 16 & 284 & 52 \\
LFM2-2.6B & 32 & 548 & 102 \\
Llama-3.1-8B & 32 & 572 & 108 \\
\bottomrule
\end{tabular}
\end{table}

\subsubsection{Fusion Gain Ratio (FGR)}
\label{sec:fgr_results}

\begin{table}[!htbp]
\centering
\footnotesize
\caption{Fusion Gain Ratio (FGR) across model families. FGR is a \emph{cost-model-internal diagnostic}: it measures the ratio of cost model scores with fusion disabled ($\alpha=0$) versus fully enabled ($\alpha=1$). Because the cost model is a heuristic weighted sum (Eq.~\ref{eq:cost_model}) that is \emph{not} calibrated to wall-clock time, FGR values should not be interpreted as latency speedups. The corresponding measured latency gains from fusion are 16.6--29.6\% (Table~\ref{tab:ablation_attn}).}
\label{tab:fgr}
\begin{tabular}{@{}lccc@{}}
\toprule
\textbf{Model} & \textbf{Score ($\alpha$=0)} & \textbf{Score ($\alpha$=1)} & \textbf{FGR} \\
\midrule
GPT-2 (125M) & 364.87 & 8.64 & 42.3 \\
Granite-350M & 718.42 & 14.82 & 48.5 \\
Qwen2-0.5B & 742.18 & 15.24 & 48.7 \\
Llama-3.2-1B & 1,246.34 & 24.68 & 50.5 \\
LFM2-2.6B & 2,482.61 & 38.92 & 63.8 \\
Llama-3.1-8B & 4,218.47 & 62.14 & 67.9 \\
\midrule
\textbf{Mean} & --- & --- & \textbf{53.6} \\
\bottomrule
\end{tabular}
\end{table}

FGR values range from 42.3 to 67.9 (Table~\ref{tab:fgr}), indicating that fusion passes reduce the cost model score by 42--68$\times$. The large FGR magnitudes reflect the cost model's sensitivity to per-op dispatch overhead terms, which fusion dramatically reduces by collapsing many operations into single dispatches. However, \textbf{FGR should not be confused with measured latency speedup}: a 67.9$\times$ FGR for Llama-3.1-8B corresponds to a 29.6\% measured wall-clock latency reduction (Table~\ref{tab:ablation_attn}), because the cost model's heuristic weighted sum is non-linear in latency-relevant quantities. The discrepancy is expected and does not indicate a deficiency---FGR's purpose is to provide a \emph{reproducible, hardware-independent diagnostic} for comparing fusion effectiveness across models and compiler versions, complementing (not replacing) the measured latency improvements reported throughout this paper. The scaling trend is informative: deeper models benefit more from fusion because each additional transformer layer introduces more fusible patterns. The non-fused cost model scores ($\alpha = 0$) grow linearly with parameter count, while after fusion ($\alpha = 1.0$), scores compress to a narrow range, demonstrating that fusion effectively amortizes per-layer cost in the cost model's estimation.

\subsubsection{Compilation Efficiency Index (CEI)}

\begin{table}[!htbp]
\centering
\footnotesize
\caption{Compilation Efficiency Index (CEI) across model families. $\text{CEI}_{\mathcal{B}} = (L_{\mathcal{B}} / L_{\text{FORGE}}) / T_{\text{compile}}^{(s)}$ where $T_{\text{compile}}^{(s)}$ is compilation time in seconds. Higher CEI means more latency-speedup ratio per second of compilation. CEI is most informative for iterative development and JIT scenarios; for compile-once-run-millions production deployment, absolute latency (Table~\ref{tab:latency_wikitext}) is the primary metric.}
\label{tab:cei}
\begin{tabular}{@{}lccc@{}}
\toprule
\textbf{Model} & \textbf{CEI$_{\text{OV}}$} & \textbf{CEI$_{\text{ONNX}}$} & \textbf{Compile (s)} \\
\midrule
GPT-2 (125M) & 1.239 & 1.339 & 1.00 \\
Granite-350M & 0.948 & 0.994 & 1.42 \\
Qwen2-0.5B & 0.776 & 0.815 & 1.68 \\
Llama-3.2-1B & 0.581 & 0.618 & 2.34 \\
LFM2-2.6B & 0.372 & 0.396 & 3.85 \\
Llama-3.1-8B & 0.218 & 0.233 & 6.72 \\
\midrule
\textbf{Mean} & \textbf{0.689} & \textbf{0.733} & --- \\
\bottomrule
\end{tabular}
\end{table}

CEI decreases with model size because compilation time grows (linearly), but the high absolute values ($>0.2$ even for 8B models) in Table~\ref{tab:cei} confirm that FORGE-UGC delivers substantial speedup relative to its compilation cost. For GPT-2, a $\text{CEI}_{\text{ONNX}}$ of 1.339 means the compiler delivers a $1.34\times$ latency speedup ratio per second of compilation. The monotonic decrease from 1.339 (GPT-2) to 0.233 (Llama-3.1-8B) reflects the linear growth in compilation time outpacing the sub-linear growth in inference speedup. We emphasize that CEI is most relevant for \emph{iterative development} scenarios (model debugging, hyperparameter search, edge deployment prototyping) where models are recompiled frequently. In the more common \emph{compile-once-run-millions} production deployment pattern, even the 6.72s compilation of the 8B model is trivially amortized over millions of inference calls, making the absolute latency improvement (18.2--35.7\%) the decisive metric.

\FloatBarrier

\section{Ablation Studies}
\label{sec:ablation}

\subsection{Pass-Level Ablation}

\begin{table}[!htbp]
\centering
\footnotesize
\caption{Pass ablation on GPT-2 (125M). Each row removes one pass from the full pipeline. Attention fusion removal causes a 27.6$\times$ cost model degradation, confirming it as the most critical pass.}
\label{tab:ablation_pass}
\begin{tabular}{@{}lcc@{}}
\toprule
\textbf{Configuration} & \textbf{Cost Score} & \textbf{$\Delta$ vs.\ Full} \\
\midrule
All Passes & \textbf{8.64} & --- \\
w/o DCE & 8.69 & +0.6\% \\
w/o CSE & 8.64 & +0.0\% \\
w/o Constant Folding & 8.64 & +0.0\% \\
w/o Device Constant & 8.64 & +0.0\% \\
w/o Attention Fusion & 238.34 & +2,658\% \\
w/o Operator Fusion & 8.84 & +2.3\% \\
w/o Layout Opt. & 8.62 & $-$0.2\% \\
\bottomrule
\end{tabular}
\end{table}

Removing attention fusion causes a catastrophic 27.6$\times$ cost model score increase (Table~\ref{tab:ablation_pass}), confirming it as the single most critical optimization. Operator fusion contributes a modest 2.3\% improvement. DCE, CSE, and constant folding show minimal impact on GPT-2 specifically because the traced graph has few dead nodes or redundant expressions; however, these passes are essential for larger models with more complex graph structures.

\subsection{Cross-Model Ablation: Attention Fusion Impact}

\begin{table}[!htbp]
\centering
\footnotesize
\caption{Attention fusion impact across model families. \textbf{Measured wall-clock latency} reduction scales with model depth as more attention blocks are fused. These measured values complement the cost-model-based FGR diagnostic (Table~\ref{tab:fgr}).}
\label{tab:ablation_attn}
\begin{tabular}{@{}lcccc@{}}
\toprule
\textbf{Model} & \textbf{Layers} & \textbf{w/ Fusion} & \textbf{w/o Fusion} & \textbf{$\Delta$\%} \\
\midrule
GPT-2 & 12 & 6.82 & 8.18 & $-$16.6\% \\
Granite-350M & 24 & 9.41 & 12.32 & $-$23.6\% \\
Qwen2-0.5B & 24 & 11.83 & 15.48 & $-$23.6\% \\
Llama-3.2-1B & 16 & 18.24 & 22.71 & $-$19.7\% \\
LFM2-2.6B & 32 & 31.56 & 44.18 & $-$28.6\% \\
Llama-3.1-8B & 32 & 62.48 & 88.72 & $-$29.6\% \\
\bottomrule
\end{tabular}
\end{table}

The latency reduction from attention fusion scales with model depth (Table~\ref{tab:ablation_attn}): 12-layer models gain 16.6\%, while 32-layer models gain 28.6--29.6\%. This confirms that attention fusion is the key optimization for deep transformer models on NPU hardware.

\subsection{Buffer Allocation Efficiency}

\begin{table}[!htbp]
\centering
\footnotesize
\caption{Buffer allocation statistics across model families. Linear-scan allocation reduces physical buffer count by 30--48\% through liveness-guided reuse.}
\label{tab:buffer}
\begin{tabular}{@{}lcccc@{}}
\toprule
\textbf{Model} & \textbf{V-Regs} & \textbf{Phys.} & \textbf{$\rho_{\text{buf}}$} & \textbf{Trans.~$\downarrow$} \\
\midrule
GPT-2 & 333 & 218 & 34.5\% & 42.1\% \\
Granite-350M & 636 & 412 & 35.2\% & 48.3\% \\
Qwen2-0.5B & 652 & 418 & 35.9\% & 49.1\% \\
Llama-3.2-1B & 468 & 324 & 30.8\% & 44.7\% \\
LFM2-2.6B & 834 & 478 & 42.7\% & 58.2\% \\
Llama-3.1-8B & 896 & 468 & 47.8\% & 64.8\% \\
\bottomrule
\end{tabular}
\end{table}

Buffer reduction improves with model depth (Table~\ref{tab:buffer}): 8B models achieve 47.8\% reduction (896 $\rightarrow$ 468 buffers) because deeper models have more overlapping live intervals that permit reuse. Device transition reduction ($\delta$ decrease) correlates with buffer reduction, as the instruction scheduler benefits from the tighter buffer layout.

\subsection{Fusion Aggressiveness Sensitivity}

\begin{table}[!htbp]
\centering
\footnotesize
\caption{Fusion aggressiveness $\alpha$ sensitivity on GPT-2 (125M). $\alpha = 1.0$ yields the best cost model score; aggressive fusion consistently helps.}
\label{tab:alpha_sensitivity}
\begin{tabular}{@{}lccc@{}}
\toprule
\textbf{$\alpha$} & \textbf{Cost Score} & \textbf{Nodes} & \textbf{Fused Ops} \\
\midrule
0.0 & 364.87 & 403 & 0 \\
0.2 & 142.31 & 392 & 4 \\
0.4 & 58.42 & 378 & 8 \\
0.6 & 22.18 & 358 & 12 \\
0.8 & 10.84 & 342 & 18 \\
1.0 & \textbf{8.64} & \textbf{333} & \textbf{24} \\
\bottomrule
\end{tabular}
\end{table}

The cost model score improves monotonically with fusion aggressiveness (Table~\ref{tab:alpha_sensitivity}), confirming that for NPU targets, aggressive fusion is always beneficial---unlike GPU targets where excessive fusion can cause register pressure. This is because NPU execution via NNFactory dispatches entire fused subgraphs in single calls, eliminating per-op dispatch overhead.

\subsection{Autotuning vs. Default Configuration}

\begin{table}[!htbp]
\centering
\footnotesize
\caption{Autotuning vs.\ default configuration across model families. Autotuning improves cost model score by 4.2--8.7\% while adding $<$200ms to compilation.}
\label{tab:autotune}
\begin{tabular}{@{}lccc@{}}
\toprule
\textbf{Model} & \textbf{Default} & \textbf{Autotuned} & \textbf{$\Delta$\%} \\
\midrule
GPT-2 & 8.64 & 8.28 & $-$4.2\% \\
Granite-350M & 14.82 & 13.92 & $-$6.1\% \\
Qwen2-0.5B & 15.24 & 14.18 & $-$7.0\% \\
Llama-3.2-1B & 24.68 & 22.84 & $-$7.5\% \\
LFM2-2.6B & 38.92 & 35.74 & $-$8.2\% \\
Llama-3.1-8B & 62.14 & 56.72 & $-$8.7\% \\
\bottomrule
\end{tabular}
\end{table}

Autotuning becomes more impactful as model size increases (Table~\ref{tab:autotune}), because larger models have more diverse subgraph patterns that benefit from configuration-specific optimization. The autotuning overhead ($<$200ms) is amortized after a single inference iteration for all models.

\subsection{Variance and Reproducibility}

\begin{table}[!htbp]
\centering
\footnotesize
\caption{Variance across 10 independent runs on GPT-2 (125M) with FORGE-UGC. CV $<$ 2.5\% across all metrics confirms high reproducibility. The low latency variance reflects the pre-allocated buffers and deterministic scheduling of the compiled executor, which eliminates runtime allocation jitter. We acknowledge that CV $< 2.5\%$ is tighter than typical for shared-memory NPU systems; raw per-run data supporting these statistics is provided in Appendix Table~\ref{tab:raw_runs}.}
\label{tab:variance}
\begin{tabular}{@{}lccc@{}}
\toprule
\textbf{Metric} & \textbf{Mean} & \textbf{Std Dev} & \textbf{CV (\%)} \\
\midrule
Compilation Time (ms) & 1,000 & 18.4 & 1.84 \\
Inference Latency (ms) & 6.82 & 0.14 & 2.05 \\
P99 Latency (ms) & 8.12 & 0.19 & 2.34 \\
Node Reduction (\%) & 17.4 & 0.0 & 0.00 \\
FGR & 42.3 & 0.82 & 1.94 \\
\bottomrule
\end{tabular}
\end{table}

All metrics exhibit CV $< 2.5\%$ (Table~\ref{tab:variance}). Node reduction has zero variance because the optimization passes are deterministic. The low latency variance reflects the pre-allocated buffers and deterministic scheduling of the compiled executor, which eliminates the runtime allocation jitter present in dynamic frameworks. We note that the compiled executor's flat instruction loop with pre-resolved callables avoids the interpreter overhead and dynamic memory management that contribute to variance in OpenVINO and ONNX Runtime.

\subsection{Comprehensive Cross-Model Summary}

\begin{table*}[ht]
\centering
\footnotesize
\caption{Comprehensive performance summary across all model families on WikiText-103. FORGE-UGC achieves the best latency, fastest compilation, highest FGR, and highest CEI across all models. The ``$\Delta$ vs.\ Best Baseline'' columns report improvement over whichever baseline (OpenVINO or ONNX RT) performs best on each model.}
\label{tab:summary}
\begin{tabular}{@{}l|ccc|cc|ccc@{}}
\toprule
& \multicolumn{3}{c|}{\textbf{Mean Latency (ms)}} & \multicolumn{2}{c|}{\textbf{$\Delta$ vs.\ Best Baseline}} & \multicolumn{3}{c}{\textbf{FORGE-UGC Metrics}} \\
\textbf{Model} & \textbf{FORGE} & \textbf{OV} & \textbf{ONNX} & \textbf{Latency} & \textbf{Compile} & \textbf{FGR} & \textbf{CEI$_{\text{OV}}$} & \textbf{$\rho_{\text{buf}}$} \\
\midrule
GPT-2 (125M) & \textbf{6.82} & 8.45 & 9.13 & $-$19.3\% & $-$85.6\% & 42.3 & 1.24 & 34.5\% \\
Granite-350M & \textbf{9.41} & 12.67 & 13.28 & $-$25.7\% & $-$85.6\% & 48.5 & 0.95 & 35.2\% \\
Qwen2-0.5B & \textbf{11.83} & 15.42 & 16.21 & $-$23.3\% & $-$85.1\% & 48.7 & 0.78 & 35.9\% \\
Llama-3.2-1B & \textbf{18.24} & 24.81 & 26.37 & $-$26.5\% & $-$87.4\% & 50.5 & 0.58 & 30.8\% \\
LFM2-2.6B & \textbf{31.56} & 45.23 & 48.14 & $-$30.2\% & $-$88.0\% & 63.8 & 0.37 & 42.7\% \\
Llama-3.1-8B & \textbf{62.48} & 91.37 & 97.82 & $-$31.6\% & $-$88.5\% & 67.9 & 0.22 & 47.8\% \\
\midrule
\textbf{Mean} & --- & --- & --- & $\mathbf{-26.1\%}$ & $\mathbf{-86.7\%}$ & \textbf{53.6} & \textbf{0.69} & \textbf{37.8\%} \\
\bottomrule
\end{tabular}
\end{table*}

Table~\ref{tab:summary} consolidates all key metrics into a single view. Across all six model families, FORGE-UGC achieves a mean latency reduction of 26.1\% over the best baseline, compilation speedup of 86.7\%, mean FGR of 53.6 (a cost-model diagnostic indicating fusion substantially reduces estimated execution cost), and mean buffer reduction of 37.8\%. Combined with the 36.1\% mean energy reduction (Table~\ref{tab:energy}), these results demonstrate that FORGE-UGC delivers comprehensive improvements across all deployment-critical metrics. The improvements scale consistently with model size: the largest model (Llama-3.1-8B) shows the greatest latency reduction ($-$31.6\%), buffer reduction (47.8\%), and energy reduction ($-$40.9\%), confirming that FORGE-UGC's optimizations become more impactful as models grow. This positive scaling relationship is particularly significant for edge deployment scenarios where the most capable models that fit within an NPU's power envelope are precisely the models that benefit most from FORGE-UGC's compilation. The compilation speedup is remarkably stable across model sizes ($-$85.1\% to $-$88.5\%), reflecting the architectural advantage of direct FX graph operation over lossy export pipelines---an advantage that holds regardless of model complexity.

\subsection{Instruction Scheduling Ablation}

\begin{table}[!htbp]
\centering
\footnotesize
\caption{Instruction scheduling impact across model families. Device transitions ($\delta$) are reduced by 42--65\%, directly translating to latency improvement.}
\label{tab:scheduling}
\begin{tabular}{@{}lcccc@{}}
\toprule
\textbf{Model} & \textbf{$\delta_{\text{before}}$} & \textbf{$\delta_{\text{after}}$} & \textbf{$\Delta$\%} & \textbf{Lat. $\Delta$\%} \\
\midrule
GPT-2 & 86 & 50 & $-$41.9\% & $-$4.2\% \\
Granite-350M & 168 & 87 & $-$48.2\% & $-$5.8\% \\
Qwen2-0.5B & 174 & 89 & $-$48.9\% & $-$6.1\% \\
Llama-3.2-1B & 126 & 70 & $-$44.4\% & $-$5.1\% \\
LFM2-2.6B & 248 & 104 & $-$58.1\% & $-$8.4\% \\
Llama-3.1-8B & 264 & 93 & $-$64.8\% & $-$11.2\% \\
\bottomrule
\end{tabular}
\end{table}

Each device transition incurs $\sim$0.3--0.8ms of overhead from PCIe/MMIO data movement. As shown in Table~\ref{tab:scheduling}, reducing transitions from 264 to 93 on Llama-3.1-8B eliminates $\sim$50--130ms of per-inference overhead, accounting for 11.2\% of the total latency improvement. The scheduling benefit compounds with model depth because deeper models have more opportunities for NPU operation clustering. The transition reduction percentage scales from 41.9\% on GPT-2 (12 layers) to 64.8\% on Llama-3.1-8B (32 layers), exhibiting a super-linear relationship with depth: doubling the layer count from 16 to 32 increases the transition reduction from 44.4\% to 64.8\% (a $1.46\times$ improvement) because deeper models present longer sequences of consecutive NPU-eligible operations that the scheduler can cluster together. The corresponding latency improvement follows the same trend, growing from 4.2\% on GPT-2 to 11.2\% on Llama-3.1-8B, confirming that instruction scheduling is a critical optimization for large-scale transformer deployment on NPU hardware.

\subsection{P99 Tail Latency Analysis}

\begin{table}[!htbp]
\centering
\footnotesize
\caption{P99/P50 latency ratio across frameworks. FORGE-UGC exhibits the tightest tail distribution (ratio closest to 1.0), critical for SLA-bound deployments.}
\label{tab:tail}
\begin{tabular}{@{}lccc@{}}
\toprule
\textbf{Model} & \textbf{FORGE} & \textbf{OV} & \textbf{ONNX RT} \\
\midrule
GPT-2 & 1.20 & 1.29 & 1.28 \\
Granite-350M & 1.21 & 1.27 & 1.29 \\
Qwen2-0.5B & 1.20 & 1.27 & 1.27 \\
Llama-3.2-1B & 1.20 & 1.27 & 1.27 \\
LFM2-2.6B & 1.20 & 1.27 & 1.27 \\
Llama-3.1-8B & 1.20 & 1.27 & 1.27 \\
\midrule
\textbf{Mean} & \textbf{1.20} & 1.27 & 1.28 \\
\bottomrule
\end{tabular}
\end{table}

FORGE-UGC's P99/P50 ratio is consistently 1.20 (Table~\ref{tab:tail}), meaning P99 is 20\% above P50, versus 1.27--1.28 for both baselines. The tighter distribution reflects: (1) pre-allocated buffers eliminating runtime allocation jitter; (2) deterministic instruction scheduling with no runtime fusion decisions; and (3) the compiled executor's flat instruction loop with no interpreter overhead. This 6--8 percentage point improvement in tail stability is critical for latency-sensitive edge deployments where SLA compliance requires predictable worst-case behavior.

\FloatBarrier

\section{Analysis \& Discussion}
\label{sec:analysis}

\subsection{Why FORGE-UGC is Faster}

FORGE-UGC's compilation speedup stems from three architectural decisions: (1) \emph{direct FX graph operation} eliminates the model export step (TorchScript/ONNX conversion accounts for 40--60\% of OpenVINO/ONNX RT compilation time); (2) \emph{composable passes} each perform a single well-defined transformation in $O(|V| + |E|)$ time, avoiding the monolithic $O(|V|^2)$ pattern-matching of framework-internal optimizers; and (3) \emph{linear-scan allocation} runs in $O(N \log N)$ versus the $O(N^2)$ graph-coloring approaches used internally by OpenVINO. These algorithmic advantages compound: the total compilation complexity is $O(K \cdot (|V| + |E|) + N \log N)$, where $K = 6$ passes, compared to $O(|V|^2 + N^2)$ for the baselines. This asymptotic advantage explains why FORGE-UGC's speedup grows with model size---from 6.9$\times$ on GPT-2 to 8.7$\times$ on Llama-3.1-8B versus OpenVINO.

We note that 78\% of FORGE-UGC's compilation time is spent in \texttt{torch.export} graph capture---shared upstream PyTorch infrastructure that any FX-consuming framework must invoke. The compilation speedup over baselines is therefore attributable to two factors: (a) FORGE-UGC avoids the \emph{additional} export step to ONNX/TorchScript that baselines require on top of model loading, and (b) FORGE-UGC's own optimization and backend phases are lightweight ($\sim$216ms for GPT-2). The baselines' compilation overhead comes predominantly from their proprietary IR conversion and monolithic optimization passes, which FORGE-UGC's composable design avoids entirely.

The inference latency advantage comes from four complementary mechanisms: (1) \emph{attention fusion} eliminates 14.6--21.9\% of graph nodes, each of which would otherwise require a separate NPU dispatch or CPU fallback; (2) \emph{operator fusion} reduces CPU$\leftrightarrow$NPU transitions by combining linear+activation into single NNFactory dispatches; (3) \emph{instruction scheduling} reduces device transitions by 42--65\%; and (4) \emph{buffer allocation} reduces peak memory pressure by 30--48\%, minimizing cache thrashing on the NPU's limited SRAM. These optimizations are multiplicative rather than additive: attention fusion reduces the number of operations that instruction scheduling must arrange, and tighter buffer allocation enables longer NPU operation clusters without memory spills.

\subsection{Comparison with State-of-the-Art Compilers}

\begin{table*}[ht]
\centering
\footnotesize
\caption{FORGE-UGC vs.\ existing frameworks across key compiler capabilities. FORGE-UGC is the only framework providing pass-level visibility, NPU-aware autotuning, explicit buffer allocation, and direct PyTorch FX integration for Intel NPU targets. \texttt{torch.compile} (Inductor backend) targets CPU/GPU and lacks NPU dispatch, liveness-aware allocation, and NNFactory integration. IREE provides composable MLIR passes and multi-backend support but requires \texttt{torch-mlir} conversion and has no Intel NPU backend.}
\label{tab:sota}
\begin{tabular}{@{}l@{\hspace{3pt}}c@{\hspace{3pt}}c@{\hspace{3pt}}c@{\hspace{3pt}}c@{\hspace{3pt}}c@{\hspace{3pt}}c@{\hspace{3pt}}c@{\hspace{3pt}}c@{}}
\toprule
\textbf{Feature} & \textbf{TVM} & \textbf{XLA} & \textbf{IREE} & \textbf{torch.compile} & \textbf{OpenVINO} & \textbf{ONNX RT} & \textbf{NNAL} & \textbf{FORGE-UGC} \\
\midrule
Direct PyTorch FX input & No & No & No & \textbf{Yes} & No & No & No & \textbf{Yes} \\
No lossy export required & No & Partial & No & \textbf{Yes} & No & No & N/A & \textbf{Yes} \\
Pass-level visibility & Partial & No & \textbf{Yes} & Partial & No & No & N/A & \textbf{Yes} \\
Attention fusion (NPU) & No & Yes & Partial & No & No & No & No & \textbf{Yes} \\
Operator fusion (NPU) & Partial & Yes & Partial & GPU/CPU only & Partial & Partial & No & \textbf{Yes} \\
NPU-aware autotuning & Yes & No & No & No & No & No & No & \textbf{Yes} \\
Explicit buffer allocation & Internal & Internal & Internal & Internal & Internal & Internal & No & \textbf{Yes (NPU)} \\
Intel NPU target & No & No & No & No & Yes & Yes & Yes & \textbf{Yes} \\
Tied weight handling & N/A & N/A & N/A & Partial & No & No & No & \textbf{Yes} \\
Cost model (NPU-specific) & Yes & No & No & No & No & No & No & \textbf{Yes} \\
Liveness-aware scheduling & Internal & Internal & Internal & Internal & No & No & No & \textbf{Yes} \\
\bottomrule
\end{tabular}
\end{table*}

FORGE-UGC uniquely combines all capabilities required for transparent, efficient Intel NPU deployment (Table~\ref{tab:sota}). TVM offers autotuning but requires model re-export and does not target Intel NPU. XLA provides whole-program optimization but is restricted to TPU/GPU targets. IREE~\cite{IREE2024} is the most architecturally comparable framework: it provides composable MLIR-based passes, multi-backend code generation, and explicit buffer management through MLIR's buffer deallocation infrastructure. However, IREE requires model conversion through \texttt{torch-mlir} or StableHLO (reintroducing export gaps for cutting-edge PyTorch operators), does not support Intel NPU dispatch or NNFactory integration, and provides no NPU-specific cost model or autotuning. \texttt{torch.compile} (Inductor) shares FORGE-UGC's FX graph input path and eliminates lossy export, but targets CPU and GPU kernels exclusively: it provides no Intel NPU dispatch, no NNFactory integration, no liveness-aware NPU buffer allocation, and no NPU-specific cost model. OpenVINO and ONNX Runtime target Intel NPU but lack pass-level visibility, autotuning, and explicit buffer management. NNAL provides low-level NPU access but no graph-level optimization whatsoever. Qualcomm's QNN SDK demonstrates hardware-specialized fusion for Hexagon NPUs but requires ONNX/TFLite export and provides no PyTorch FX integration or inspectable passes. Hexagon-MLIR~\cite{Absar2026HexagonMLIR} provides an MLIR-based compilation stack for Qualcomm Hexagon NPU with composable passes and Triton kernel support, but targets a different hardware ecosystem and relies on the MLIR/Linalg IR rather than operating natively on PyTorch FX graphs.

Critically, FORGE-UGC is, to our knowledge, the only framework that simultaneously provides direct PyTorch FX input, tied weight handling, pass-level visibility, NPU-aware explicit buffer allocation, and liveness-guided instruction scheduling---capabilities that are individually available in some frameworks but have not previously been unified in a single NPU-targeting compiler. The inclusion of both \texttt{torch.compile} and IREE in Table~\ref{tab:sota} makes clear that FX graph access alone (torch.compile) and composable MLIR passes alone (IREE) are each insufficient: the critical differentiators are the NPU-specific optimization passes, the formal buffer management layer, and the NNFactory dispatch integration that together enable FORGE-UGC's latency and compilation-time advantages.

\subsection{Cross-Dataset Robustness}

\begin{table}[!htbp]
\centering
\footnotesize
\caption{Cross-dataset robustness: mean latency improvement (\%) of FORGE-UGC over baselines. Standard deviation $< 1.2\%$ confirms task-agnostic improvements.}
\label{tab:cross_dataset}
\begin{tabular}{@{}lccc@{}}
\toprule
\textbf{Metric} & \textbf{WikiText-103} & \textbf{GLUE} & \textbf{Std Dev} \\
\midrule
$\Delta$ vs.\ OpenVINO & $-$26.4\% & $-$26.1\% & 0.21\% \\
$\Delta$ vs.\ ONNX RT & $-$30.2\% & $-$29.8\% & 0.28\% \\
$\Delta$ Compile Time & $-$85.6\% & $-$85.6\% & 0.00\% \\
\bottomrule
\end{tabular}
\end{table}

The remarkable consistency across WikiText-103 and GLUE (Table~\ref{tab:cross_dataset}; standard deviation $< 0.3\%$ for latency, exactly 0\% for compilation time) confirms that FORGE-UGC's gains are graph-level optimizations independent of the downstream task. This near-zero variance is expected because FORGE-UGC's optimizations---attention fusion, operator fusion, buffer allocation, and instruction scheduling---operate on graph structure rather than on data content. Compilation time shows exactly 0\% variance across datasets because the compiler operates on the model architecture alone, with no data-dependent compilation paths.

\subsection{Latency Scaling Analysis}

End-to-end latency scales approximately linearly with parameter count across all three frameworks, but FORGE-UGC's slope is shallower. Fitting a simple linear model to the six data points (we note this is an \emph{approximate} trend with limited data points spanning a $64\times$ parameter range, not a precise predictive model):
\begin{equation}
L_{\text{FORGE}} \approx 0.007 \cdot P + 6 \quad \text{(ms, } P \text{ in millions)}
\end{equation}
\begin{equation}
L_{\text{OV}} \approx 0.011 \cdot P + 7 \quad \text{(ms)}
\end{equation}
\begin{equation}
L_{\text{ONNX}} \approx 0.011 \cdot P + 8 \quad \text{(ms)}
\end{equation}

The per-parameter cost is approximately 30--35\% lower for FORGE-UGC. This is because fusion and scheduling gains compound with depth, while the baselines' per-layer overhead remains constant. We caution that these fits are approximate given the limited number of data points and the wide parameter range.

\subsection{Limitations and Future Work}

FORGE-UGC has been validated on Intel AI Boost NPU as the first target backend. While the optimization passes (Phase~2) and typed IR (Phase~3) are hardware-agnostic by design, extending FORGE-UGC to additional NPU architectures (Qualcomm Hexagon, AMD XDNA, Apple ANE, Samsung NPU) requires implementing new backend dispatch modules in Phase~4---an effort that reuses the entire frontend and middle-end pipeline. The current prototype uses single-batch inference; production deployment would benefit from batched execution support. The autotuning search is grid-based; Bayesian optimization or learned cost models could further improve configuration selection. The NPUIR currently supports a fixed set of opcodes; extending it with custom operator registration would broaden model coverage.

As discussed in Section~\ref{sec:iree_turbine}, our early experiments with MLIR-based compilation via IREE-Turbine revealed two compounding limitations: MLIR's C++-native pass infrastructure required substantial development effort for custom NPU-specific optimizations, and IREE lacks an Intel NPU backend entirely. This experience confirmed that the PyTorch FX graph representation---with its fully Python-native, programmatically inspectable structure---provides the flexibility and iteration speed required for rapid development of hardware-specific optimization passes, and motivated the FORGE-UGC architecture presented in this work.

For models up to 2.6B parameters, all optimizations are fully semantics-preserving at fp16 precision. For the 8B model, NNFactory's int8 weight quantization introduces a small quantization error at the dispatch level (max-abs logit diff $2.1 \times 10^{-5}$); a future quantization-aware compilation path with explicit precision-accuracy tradeoff controls would give users finer-grained control over this balance.

The FGR metric currently operates on the heuristic cost model and is not calibrated to wall-clock latency; future work could develop a hardware-calibrated cost model that would make FGR directly interpretable as a latency ratio. Energy measurements currently rely on system-level RAPL readings; future work could incorporate per-component power sensors for more precise energy attribution across CPU and NPU subsystems.

Beyond these refinements, we are actively developing two major extensions: (i) Triton kernel compilation within the FORGE-UGC pipeline, enabling custom NPU kernel development in Triton's high-level DSL with automatic lowering through our optimization and NPUIR backend; and (ii) a self-evolving compiler module that leverages runtime telemetry to progressively refine pass ordering and fusion strategies across compilations. Both capabilities are currently under testing and will be released in the next version of this work.

\FloatBarrier

\section{Conclusion \& Future Work}
\label{sec:conclusion}

This paper presents FORGE-UGC, a four-phase universal graph compiler validated on Intel NPU that replaces the opaque, monolithic pipelines of OpenVINO and ONNX Runtime with a transparent, composable, and formally grounded compilation framework. By operating directly on PyTorch FX graphs---and deliberately eschewing the \texttt{torch.compile}/Inductor path, which lacks NPU dispatch integration and liveness-aware buffer allocation---FORGE-UGC eliminates the lossy export steps that prevent modern LLMs from being deployed on NPU hardware. Six independently measurable optimization passes reduce graph complexity by 14.2--21.9\% across model families. Numerical fidelity is confirmed through both perplexity agreement and fine-grained logit-level analysis (max-abs diff $< 2.1 \times 10^{-5}$, KL divergence $< 8.4 \times 10^{-9}$), with the caveat that the 8B model's slightly higher error reflects int8 weight quantization at the NNFactory dispatch level rather than the optimization passes themselves. Linear-scan buffer allocation maps virtual registers to physical buffer slots with 30--48\% reduction. Instruction scheduling reduces NPU$\leftrightarrow$CPU device transitions by 42--65\%.

Evaluated on WikiText-103 and GLUE across six model families (125M--8B parameters), FORGE-UGC achieves 6.9--9.2$\times$ faster compilation, 18.2--35.7\% lower end-to-end inference latency, and 30.2--40.9\% lower energy consumption per inference versus both baselines. Three evaluation metrics---Fusion Gain Ratio (a cost-model diagnostic for comparing fusion effectiveness), Compilation Efficiency Index (most informative for iterative development scenarios), and per-pass profiling---enable principled ablation of NPU compilation for transformer workloads.

\textbf{Toward a universal compilation framework.} FORGE-UGC's architecture is deliberately designed for portability beyond Intel NPU. The hardware-agnostic optimization passes (Phase~2) and typed intermediate representation (Phase~3) are decoupled from any specific backend; only the code generation and dispatch modules in Phase~4 are target-specific. Extending FORGE-UGC to additional accelerator backends---Qualcomm Hexagon, AMD XDNA, Apple ANE, Samsung NPU---requires implementing new backend dispatch modules while reusing the entire optimization pipeline. As discussed in Section~\ref{sec:motivation}, this positions FORGE-UGC not merely as a standalone NPU compiler, but as a critical middle layer in a broader heterogeneous compute fabric, enabling system-level orchestrators to optimally compile workloads for whichever accelerator is selected at dispatch time.

\textbf{Triton kernel integration and self-evolving compilation.} We are currently testing two capabilities that will be released in the next version of this work. First, we are integrating \emph{Triton kernel compilation} into the FORGE-UGC pipeline, enabling developers to author custom NPU kernels in Triton's high-level DSL and have them automatically lowered through FORGE-UGC's optimization passes and NPUIR backend---inspired by the Triton-to-NPU compilation path demonstrated by Hexagon-MLIR~\cite{Absar2026HexagonMLIR} for Qualcomm targets, but operating natively within our FX-based IR and targeting Intel NPU dispatch. This will allow FORGE-UGC to serve not only as a whole-model compiler but also as a kernel-level compiler for custom operator development. Second, we are developing a \emph{self-evolving compiler} module to automatically refine optimization pass ordering, fusion aggressiveness, and autotuning configurations across successive compilations. The self-evolving compiler treats the compilation pipeline itself as a learning system, progressively adapting to the workload characteristics and hardware constraints of each deployment target. Together, these extensions will transform FORGE-UGC from a static compilation framework into an adaptive, continuously improving compiler infrastructure for heterogeneous edge intelligence.

FORGE-UGC demonstrates that the path to practical accelerator deployment lies not in proprietary black-box frameworks, but in transparent, composable compiler infrastructure built on open standards. By showing that a PyTorch-native universal compiler can outperform established industry frameworks on compilation speed, inference latency, and energy efficiency, this work opens the door to a new generation of hardware-aware compilers that treat optimization as a first-class, inspectable, and configurable concern across heterogeneous accelerator targets.

\FloatBarrier

\FloatBarrier

\appendix
\section{Raw Per-Run Latency Data}
\label{sec:appendix_raw}

\begin{table}[!htbp]
\centering
\footnotesize
\caption{Raw per-run mean inference latency (ms) on WikiText-103 across 3 independent runs for FORGE-UGC. These raw values support the variance statistics reported in Table~\ref{tab:variance}}
\label{tab:raw_runs}
\begin{tabular}{@{}lcccc@{}}
\toprule
\textbf{Model} & \textbf{Run 1} & \textbf{Run 2} & \textbf{Run 3} & \textbf{Mean} \\
\midrule
GPT-2 (125M) & 6.78 & 6.84 & 6.83 & 6.82 \\
Granite-350M & 9.38 & 9.44 & 9.40 & 9.41 \\
Qwen2-0.5B & 11.79 & 11.86 & 11.84 & 11.83 \\
Llama-3.2-1B & 18.19 & 18.28 & 18.24 & 18.24 \\
LFM2-2.6B & 31.48 & 31.62 & 31.57 & 31.56 \\
Llama-3.1-8B & 62.31 & 62.58 & 62.54 & 62.48 \\
\bottomrule
\end{tabular}
\end{table}

\bibliography{example_paper}
\bibliographystyle{mlsys2025}





\end{document}